# Simulation of Turbulent Flow around a Generic High-Speed Train using Hybrid Models of RANS Numerical Method with Machine Learning

Alireza Hajipour[1], Arash Mirabdolah Lavasani[1], Mohammad Eftekhari Yazdi[1], Amir Mosavi[2,3], Shahaboddin Shamshirband[4,5*], Kwok-Wing Chau[6]

[1]Department of Mechanical Engineering, Central Tehran Branch, Islamic Azad University, Tehran, Iran (email: alirezahajipour.eng@iauctb.ac.ir, arashlavasani@iauctb.ac.ir) moh.eftekhari-yazdi@iauctb.ac.ir)

[2]Institute of Structural Mechanics, Bauhaus University Weimar, 99423 Weimar, Germany; a.mosavi@brookes.ac.uk

[3]Institute of Automation, Kalman Kando Faculty of Electrical Engineering, Obuda University, 1034 Budapest, Hungary

[4]Department for Management of Science and Technology Development, Ton Duc Thang University, Ho Chi Minh City, Vietnam

[5]Faculty of Information Technology, Ton Duc Thang University, Ho Chi Minh City, Vietnam

[6]Department of Civil and Environmental Engineering, Hong Kong Polytechnic University, Hung Hom, Hong Kong, China; (email: dr.kwok-wing.chau@polyu.edu.hk)

*Corresponding author, Email: shahaboddin.shamshirband@tdtu.edu.vn

## Abstract

In the present paper, an aerodynamic investigation of a high-speed train is performed. In the first section of this article, a generic high-speed train against a turbulent flow is simulated, numerically. The Reynolds-Averaged Navier-Stokes (RANS) equations combined with the $k$-$\omega$ SST turbulence model are applied to solve incompressible turbulent flow around a high-speed train. Flow structure, velocity and pressure contours and streamlines at some typical wind directions of $\theta$ = 0˚, 30˚, 45˚ and 60, are the most important results of this simulation. The maximum and minimum values are specified and discussed. Also, the pressure coefficient for some critical points on the train surface is evaluated. In the following, the wind direction influence the aerodynamic key parameters as drag, lift, and side forces at the mentioned wind directions are analyzed and compared. Moreover, the effects of velocity changes (50, 60, 70, 80 and 90 m/s) are estimated and compared on the above flow and aerodynamic parameters. In the second section of the paper, various data-driven methods including Gene Expression Programming (GEP), Gaussian Process Regression (GPR), and random forest (RF), are applied for predicting output parameters. So, drag, lift and side forces and also minimum and maximum of pressure coefficients for mentioned wind directions ($\theta$ = 0° to $\theta$ = 60°) and velocity (50 m/s to 90 m/s) are predicted and compared using statistical parameters. Obtained results indicated that RF in all coefficients of wind direction and most coefficients of free stream velocity provided the most accurate predictions. As a conclusion, RF may be recommended for the prediction of aerodynamic coefficients.

## 1. Introduction

The research on the flow around high-speed trains is considered a popular subject among engineering applications. Many countries have built high-speed train lines to link cities, including Germany, China, Austria, Belgium, France, Poland, Italy, Portugal, Japan, Russia, South Korea, Spain, Sweden, Taiwan, Turkey, United Kingdom, United States and etc. Nowadays, trains move at high speed, study of the aerodynamic characteristics of the air flow around them is an interesting topic. In recent years, many simulations of flow over high-speed trains have been performed. The most significand and related of them reviewed in previous research of the authors (Rashidi, 2019) which are listed as follow:

A comparison between numerical and experimental simulation of air flow over a high-speed train was investigated (Paradot, 2002). A SNCF (French high-speed train) with high Reynolds number (almost $10^9$) was used for aerodynamic analysis.

Effect of cross-wind on a high-speed train was investigated, numerically (Khier, 2002). In this paper, the flow over the German InterRegio high-speed train were simulated using Reynolds-averaged Navier-Stokes and $k$-ε turbulence model.

Effects of crosswind on the German InterRegio high-speed train was estimated (Fauchier, 2002). For the numerical simulation, the RANS approach along with RNG $k$-ε turbulent model was implemented. In this analysis, at first, a preliminary study has been done on a simple geometry of train via CFD. Then, a three-dimensional model of the train in three following cases have been investigated.

A research of aerodynamic flow characteristics on a high speed train was done (Shin, 2003). In this numerical study, the variation of in the aerodynamic forces were performed for the high speed train at the entrance of a tunnel.

A numerically investigation of the fluid flow around a high-speed train was done (Tian, 2009). As the high speed at modern train is effective in aerodynamic drag, reducing drag and consequently reducing energy consumption is one of the main issues in the development of the rail industry.

An aerodynamic characteristic of a Chinese high-speed train using RANS numerical method was performed (Zhao, 2009). In this research, the aerodynamic influences were studied with the train at the entrance and exit of a tunnel.

An optimization of aerodynamic characteristics of high-speed train using numerical method was investigated (Krajnović, 2009). By RANS numerical method, the simple polynomial response surfaces instead of complex NS simulations and Genetic Algorithm (GA) an optimization aerodynamic properties have been conducted on Swedish high-speed train X2000.

A simulation of a high-speed train at a tunnel entrance was presented (Li, 2011). For this purpose, the RANS numerical method and *k*-ε turbulent model for a viscous compressible fluid was applied.

A numerical simulation of EMU high-speed train using RANS method and RNG *k*-ε turbulent model was conducted (Wang, 2012). The main reason for this research was the pressure changes and aerodynamic forces with two trains passing alongside each other in a tunnel.

A numerical aerodynamic characteristics of a high-speed train against a crosswind using unsteady three dimensional RANS approach and *k*-ε turbulent model was done (Asress, 2014). In the simulation, two scenarios for ground as static and moving for yaw angles ranges from 30˚ to 60˚ were considered.

A numerical simulation about wind effect on a high-speed train was done (Peng, 2014). A three dimensional incompressible RANS method and *k*-ε turbulent model has been done For simulation of air flow passing a simple high-speed train.

An optimization of aerodynamic parameters for high-speed trains was investigated useing numerical method (Shuanbao, 2014). The complicated wake flow deeply affects the movement of the trains.

An aerodynamic simulation of two high-speed trains in a tunnel using RANS numerical method and RNG *k*–ε turbulence model (Chu, 2014). In this paper, a three-dimensional, compressible, turbulence model was applied to find the pressure wave.

An aerodynamic analysis of a high-speed train via numerical method was performed (Zhang, 2015). Effects of the slope angles and cut depth on the flow structure around the train were the most important goals of this article to determine. Also, the surface pressure and aerodynamic forces of train were analyzed using RANS approach combined with the eddy viscosity hypothesis in turbulence model.

A comparison between Reynolds-averaged Navier–Stokes (RANS) turbulence models and experimental wind tunnel findings by Detached Eddy Simulation (DES) for pressure on a high-speed train was conducted (Morden, 2015). The $k$-ε, $k$-ε re-normalization group (RNG), realizable $k$-ε, Spalart–Allmaras (S-A) and shear stress transport (SST $k$-ω) are the five numerical models of RANS used in this article.

An aerodynamic investigation on a simple model of high-speed train under crosswind using the numerical method of large eddy simulation (LES) was performed (Zhuang, 2015). In this study, the effects of flow diversion were investigated with two angles of $\varphi = 30^\circ$ and $60^\circ$ using LES.

A CFD simulation using RANS numerical method for a high-speed train against a crosswind for two conditions; stationary and moving was conducted (Catanzaro, 2016). The effect of each conditions was analyzed.

An aerodynamic design on a high-speed train was done (Ding, 2016). Due to the speed improvements of the high-speed trains and increasing the aerodynamic effect of the mechanical one, the effects and issues of aerodynamics were considered as the main challenge of this paper.

An aerodynamic performance of a train under crosswinds condition using RANS numerical method was investigated (Liu, 2016). In this research, via the computational fluid dynamics, the aerodynamic characteristics was done for the train on a special slope and crosswind conditions.

An aerodynamic comparison between two trains, a static and a moving was done (Premoli, 2016). In this research, using the numerical RANS method the simulation results of the relative movements of the trains between the vehicle and its infrastructure that was effective on the aerodynamic coefficients were compared.

In the present study, the key flow parameters such as pressure, velocity and the aerodynamic parameters of a turbulent air flow around a generic high-speed train are analyzed. Moreover, the mentioned parameters are predicted by the data driven methods of GEP, GPR and RF. Then, using evaluation parameters, the most accurate model is suggested.

## 2. Computational Simulation

### 2.1. Geometry Description

The train model used in present paper is a generic high-speed train one which has been used in many research studies on high-speed trains. Figure 1 shows that the geometry of the generic high-speed train model with different views. The geometric characteristics of the train are as follows:

As Figure 1, the nose form of the train is elliptical. Also, the length, height and width of the train are $7H$, $H$ and $H$, respectively ($H$ = 0.56 m).

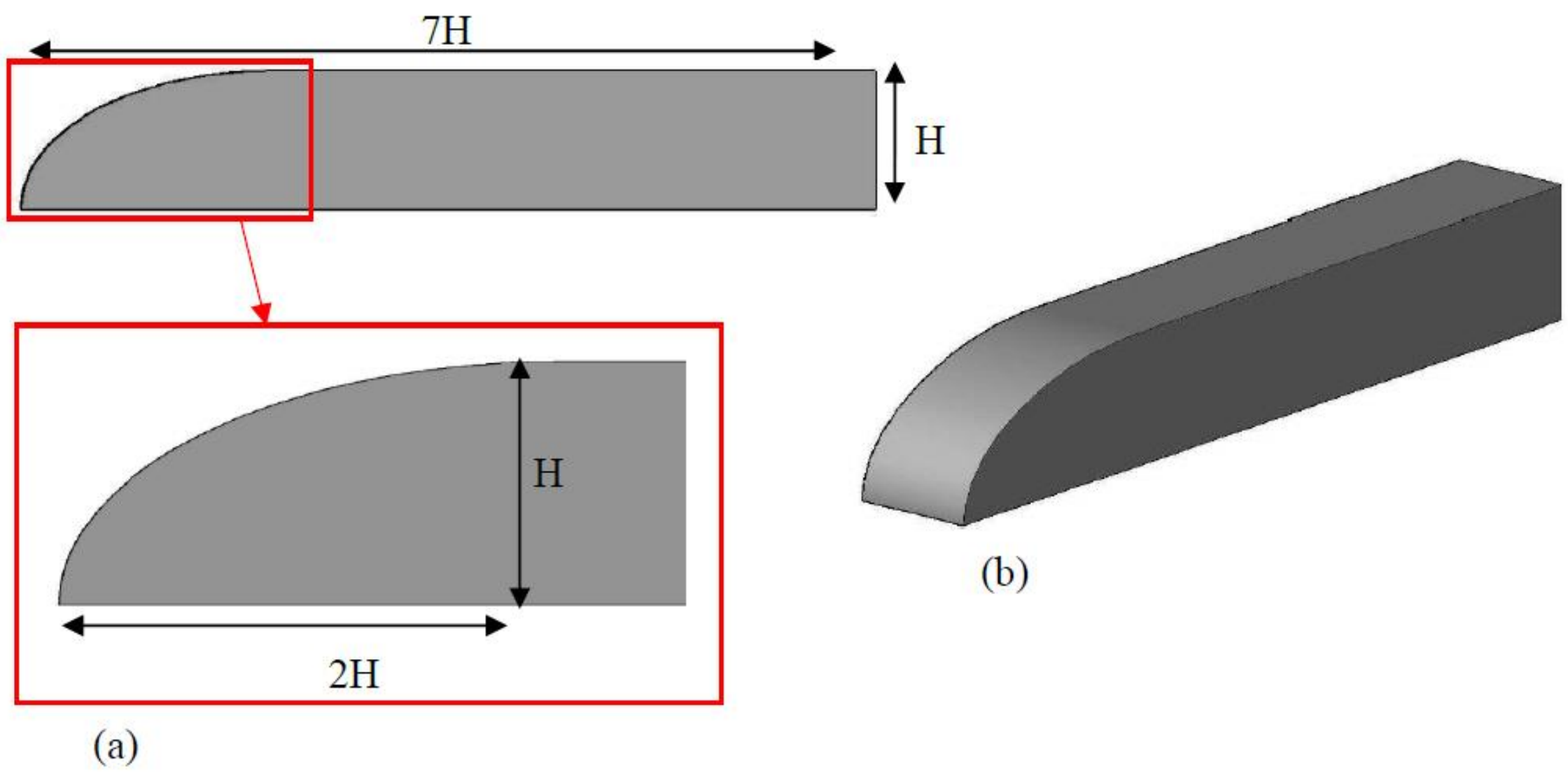

Figure 1. Geometry of the generic high-speed train; (a): Side view, (b): Isometric view.

## 2.2. Domain Description

The train model is placed 0.15*H* above the ground. The length, width, and height of the computing domain are 36*H* × 21*H* × 11.15*H*, respectively. The distance between then inlet of the domain and  the nose of the train and between the outlet of the domain and the back of the train are 8*H* and 21*H*, respectively. Moreover, the distance between train and two sides of the domain is 10*H* (see Figures 2-4).

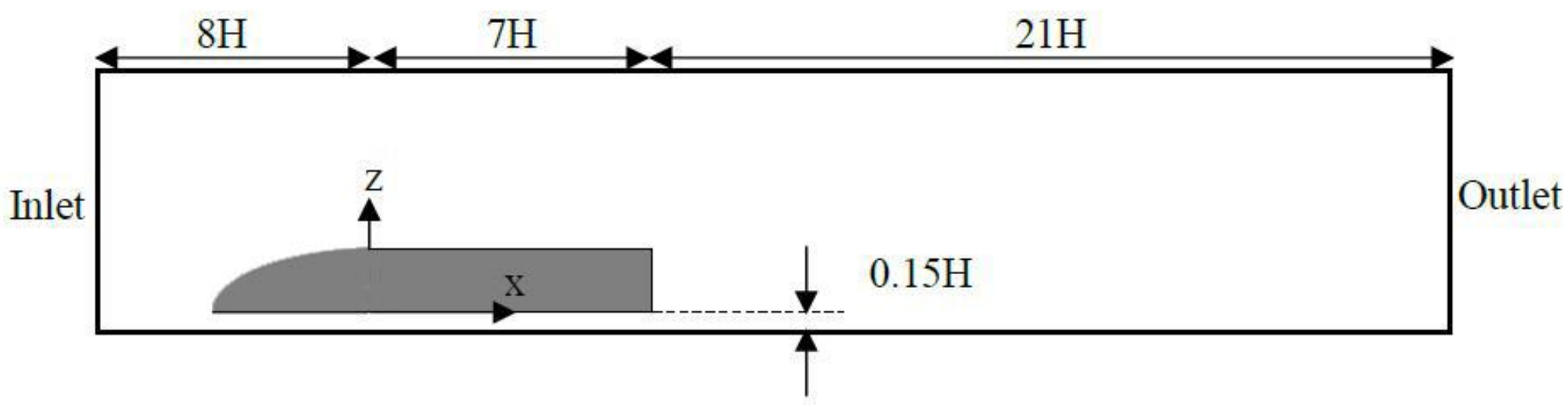

Figure 2. Side view of the computational domain.

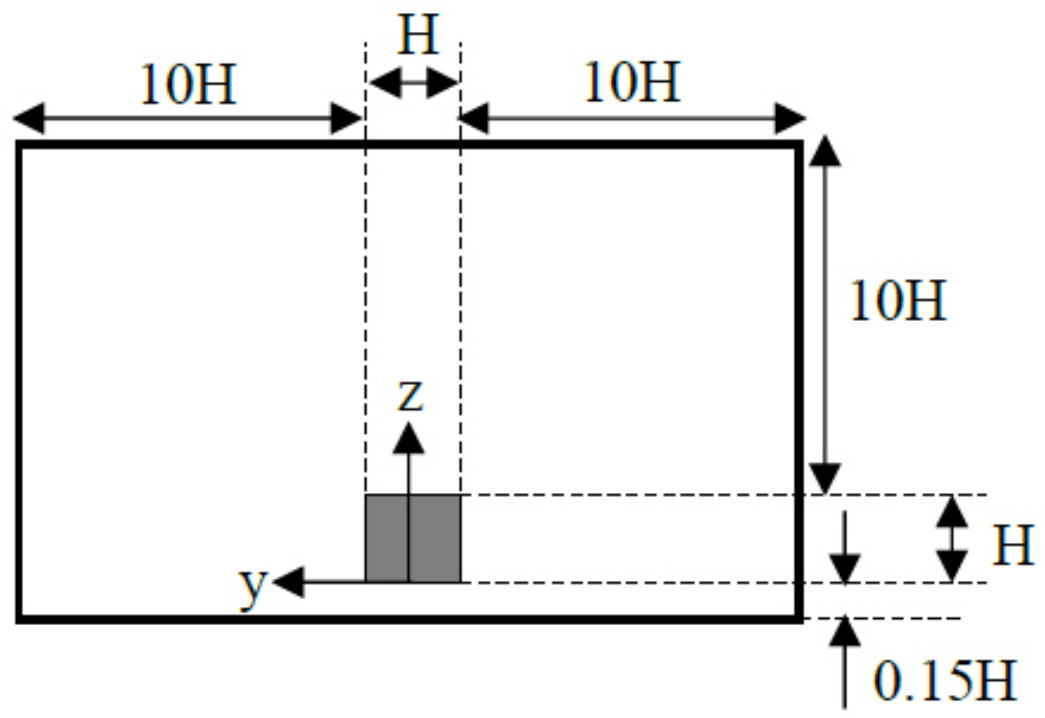

Figure 3. Front view of the computational domain.

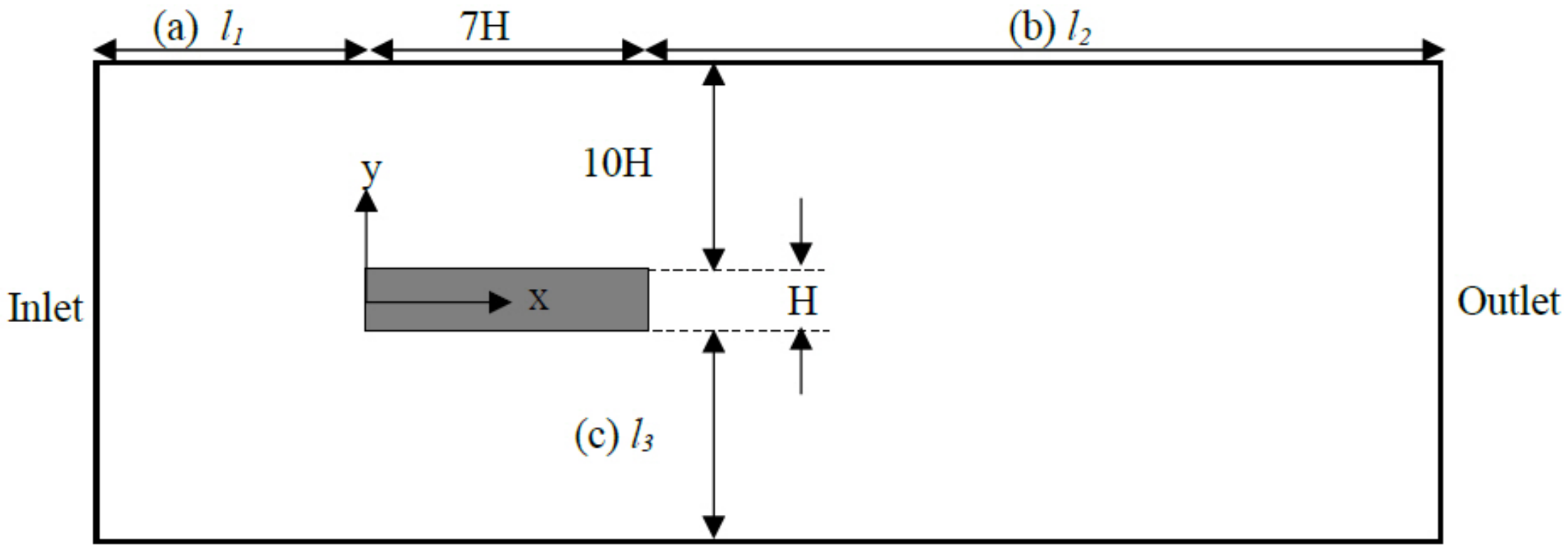

Figure 4. Top view of the computational domain.

## 2.3. Mesh Description

The used mesh in the computational domain for the different cases (wind directions of $\theta = 0^\circ$, $30^\circ$, $45^\circ$ and $60^\circ$) are 4,000,000 nodes, approximately and the $y^+$ range are between 73.2 and 94.3. For these cases, the $y^+$ must be located between 30 to 300. In the following, two refinement boxes near and around the train for more accurate analysis are used (as Figure 5).

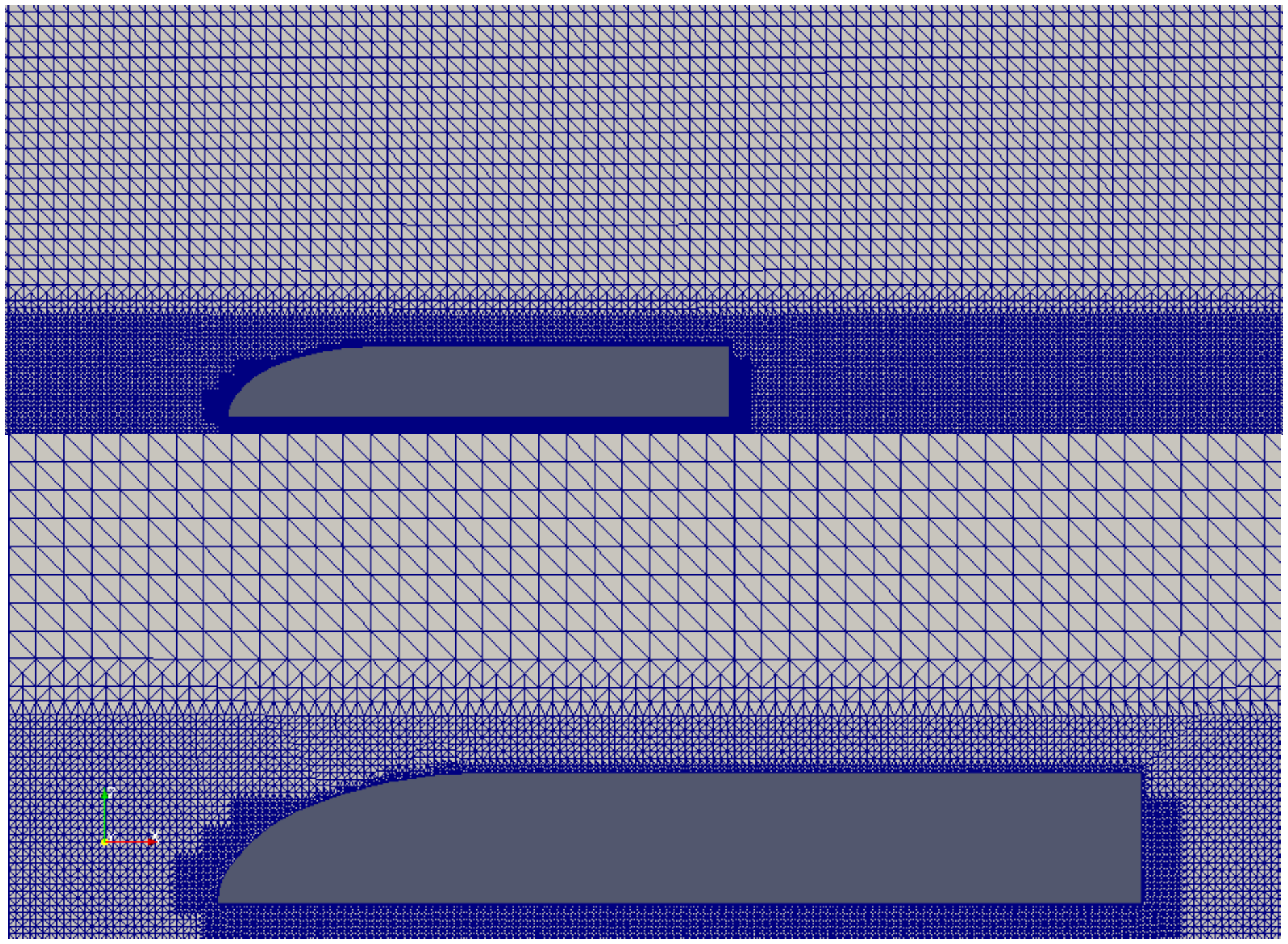

Figure 5. Wide and close view of the two refinement boxes near and around the train.

### 2.4. Boundary Conditions

The defined boundary conditions of the case are as follows:

Inlet: a uniform velocity, that represents the free stream velocity, $U_\infty$ in the x direction.

Outlet: the patch type boundary condition with a free stream value.

Sides and top of the domain: the patch type boundary condition with a free stream value.

Ground: The wall boundary condition used for the ground.

Train surface: The wall boundary condition used for the train.

Also, the Reynolds number, *Re*, according to the height of the train, $H = 0.56$ m, free stream velocity, $U_\infty = 70$ m/s, and kinematic viscosity, $v = 1.5 \times 10^{-5}$, ($Re = U_\infty \times H/ v$) is $2.6 \times 10^6$.

### 2.5. Solution Approach and Governing Equations

The air flow field around the high-speed train which defined as a 3D incompressible turbulent flow is solved by the Reynolds-Averaged Navier-Stokes (RANS) equation combined with the $k$-$\omega$ Shear-Stress Transport (SST) turbulence approach. The Reynolds-Averaged Navier-Stokes, is a time-average method of fluid flow description. In this method, instantaneous quantities are replaced by average and oscillating ones.

According to the selected solution method, The continuity and Navier-Stokes equations for the incompressible air flow around the train as follows:

$$\frac{\partial u_i}{\partial x_j} = 0 \tag{1}$$

$$\rho \frac{\partial u_i}{\partial x_j} + \rho U_j \frac{\partial u_i}{\partial x_j} = -\frac{\partial p_i}{\partial x_j} + \frac{\partial}{\partial x_j}\left[\mu\left(\frac{\partial u_j}{\partial x_j} + \frac{\partial u_i}{\partial x_j}\right)\right] \tag{2}$$

where, i, j and k = 1, 2 and 3 are related to the streamwise –x, cross-stream –y and –z direction, respectively. The velocity ingredients, $u_i$ and the pressure, $p_i$ are both nonlinear terms. Hence, there aren't any analytical solves for the problem with optional boundary conditions. The transience of the flow parameters (i.e. velocity and pressure) are divided into mean value and fluctuations as follows:

$$u_i = U_i + u'_i \tag{3}$$

$$p_i = P_i + p'_i \tag{4}$$

where $U_i$ and $P_i$ are the time-averaged, while $u'_i$ and $p'_i$ are the fluctuation terms of velocity and pressure, respectively. Substituting the Reynolds divided velocities and pressures into the Continuity and Navier-Stokes equations yields the RANS equation of motions as illustrated below:

$$\frac{\partial U_i}{\partial x_j} = 0 \tag{5}$$

$$\frac{\partial U_i}{\partial t} + U_j \frac{\partial U_i}{\partial x_j} = -\frac{1}{\rho}\frac{\partial P_i}{\partial x_i} + \frac{\partial}{\partial x_j}\left(\mu \frac{\partial U_i}{\partial x_j} - \rho \overline{u_i' u_j'}\right) \tag{6}$$

According to the Boussinesq, the Reynolds-stress tensor could be connected to the mean rate of deformation. The concept applied for the turbulence model is as below:

$$-\rho \overline{u_i' u_j'} = \mu_t \left(\frac{\partial U_i}{\partial x_j} + \frac{\partial U_j}{\partial x_i} - \frac{2}{3}\frac{\partial U_k}{\partial x_k}\delta_{ij}\right) - \frac{2}{3}\rho k \delta_{ij} \tag{7}$$

where, the turbulent kinetic energy ($k$) and the specific dissipation rate ($\omega$) are solved via the following equations:

Turbulence Kinetic Energy, $k$

$$\frac{\partial k}{\partial t} + U_j \frac{\partial k}{\partial x_j} = P_k - \beta^* k\omega + \frac{\partial}{\partial x_j}\left[(\upsilon + \sigma_k \upsilon_T)\frac{\partial k}{\partial x_j}\right] \tag{8}$$

Specific Dissipation Rate, $\omega$

$$\frac{\partial \omega}{\partial t} + U_j \frac{\partial \omega}{\partial x_j} = \alpha S^2 - \beta \omega^2 + \frac{\partial}{\partial x_j}\left[(\upsilon + \sigma_\omega \upsilon_T)\frac{\partial \omega}{\partial x_j}\right] + 2(1 - F_1)\sigma_{\omega 2}\frac{1}{\omega}\frac{\partial k}{\partial x_i}\frac{\partial \omega}{\partial x_i} \tag{9}$$

where, $\nu_T$ is the kinematic eddy viscosity which defined as follows:

$$\upsilon_T \frac{\alpha_1 k}{max(\alpha_1 \omega, S F_2)} \tag{10}$$

The following closure coefficient is applied in the paper:

$$F_2 = tanh\left[\left[max\left(\frac{2\sqrt{k}}{\beta^* \omega y}, \frac{500\upsilon}{y^2 \omega}\right)\right]^2\right] \tag{11}$$

where y is the distance to the next surface,

$$P_k = min\left(\tau_{ij}\frac{\partial U_i}{\partial x_j}, 10\beta^* k\omega\right) \tag{12}$$

$$F_1 = tanh\left\{\left\{min\left[max\left(\frac{\sqrt{k}}{\beta^* \omega y}, \frac{500\upsilon}{y^2 \omega}\right), \frac{4\sigma_{\omega 2} k}{CD_{k\omega} y^2}\right]\right\}^4\right\} \tag{13}$$

$$CD_{k\omega} = max\left(2\rho\sigma_{\omega 2}\frac{1}{\omega}\frac{\partial k}{\partial x_i}\frac{\partial \omega}{\partial x_i}, 10^{-10}\right) \tag{14}$$

$$\phi = \phi_1 F_1 + \phi_2(1 - F_1) \tag{15}$$

$$\alpha_1 = \frac{5}{9}, \qquad \alpha_2 = 0.44 \tag{16}$$

$$\beta_1 = \frac{3}{40}, \qquad \beta_2 = 0.0828, \qquad \beta^* = \frac{9}{100} \tag{17}$$

$$\sigma_{k1} = 0.85, \sigma_{k2} = 1, \sigma_{\omega 1} = 0.5, \sigma_{\omega 1} = 0.856, \tag{18}$$

## 3. Aerodynamic Results and Discussions

### 3.1. Flow Structure

It is clear that the wind direction has significant influence on the flow structure around the generic high-speed train. In this section, the influence of free stream direction at four different angles of wind directions of $\theta$ = 0˚, 30˚, 45˚ and 60˚ perpendicular to the x-axis with constant velocity magnitude is presented graphically in Figure 6. It should be noted that the obtained flow structure is predicted by SST transitional model as a time-averaged flow.

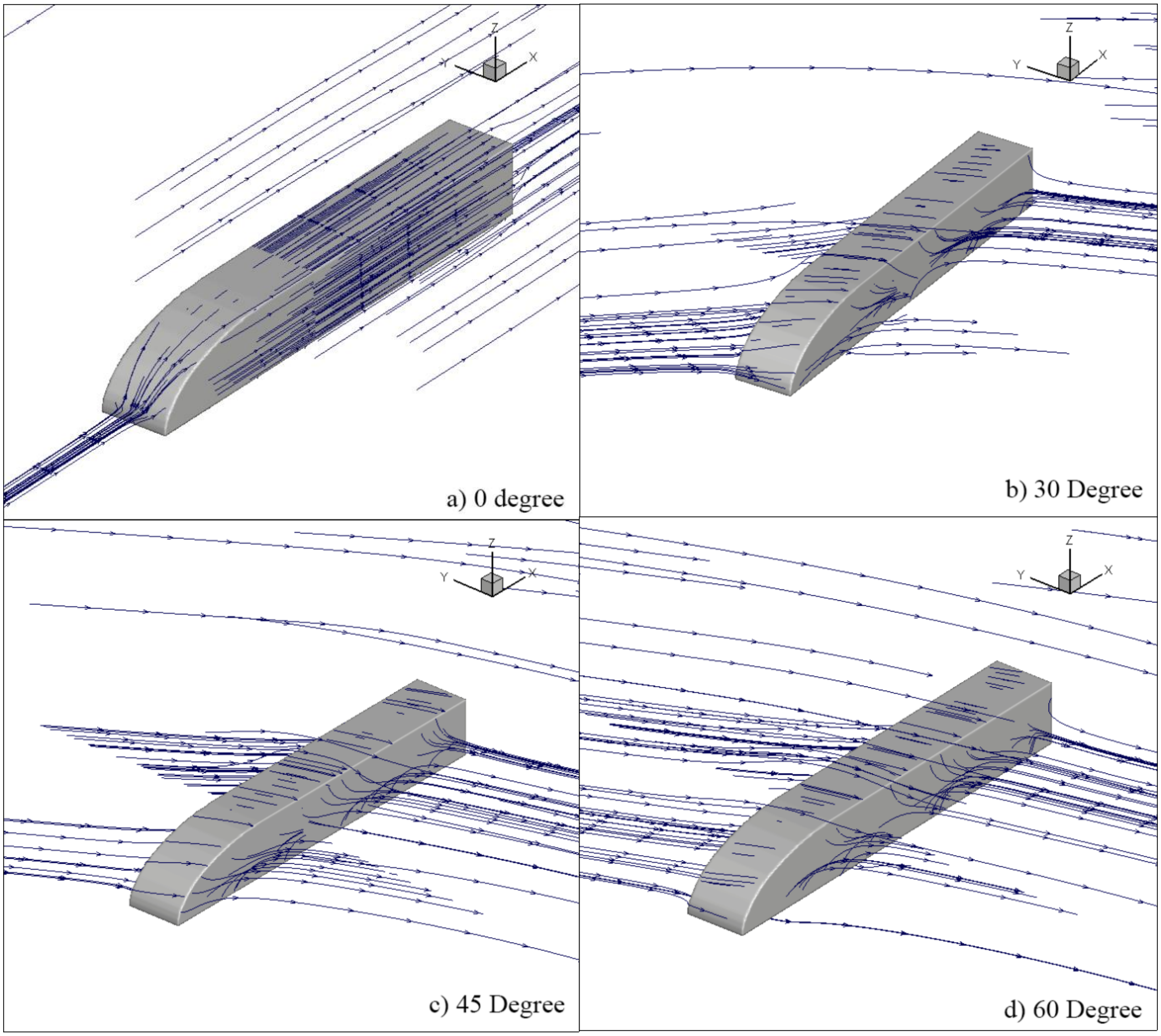

Figure 6. Three-dimensional time-averaged flow structures for different wind direction angles.

As it can be seen in Figure 10, the circulations at the lee-side of the generic train model were created. These circulations at each section along the length of model are consists of two main circulations which are kindled form roof-side and bottom-side of the model. Furthermore, the length and width of the circulations changes along the length of model. It is obvious in the figures that the nose of the model has considerable influence on the three-dimensional flow structure. The nose of model causes vortex which is inclined to positive direction of x-axis. It should be noted that the circulation existing at the leeward of the model enhances the pressure coefficient and resulted drag force.

In order to comprehensive investigation of the flow structure, the flow pattern at two most important regions were investigated for windside and leeward. The leeward flow patterns are identified by three-dimensional and two dimensional flow patterns. Since the three-dimensional flow structures at the windside will not render clear insight, the stream lines at the surface of the generic train model are obtained in Figure 7. It can be observed that the positions of the stagnation lines at the surface of model's body are almost similar for all cases. On the other hand, the stagnation lines at the model's nose are deflected to roofward of the model. It is due to the fact that the air flow is imposed to be passed form roofward due to limited space at the bottom of the train model. Totally, the variation of wind direction has no pronounced influence on the flow pattern at the windside region.

The two-dimensional total pressure contours over the train for different wind directions ($\theta$ = 0˚, 30˚, 45˚ and 60˚) are presented in Figure 8. The pressure distribution around the train are the same for the different wind directions, generally. The pressure at the front and the back of the train has the maximum and minimum values, respectively.

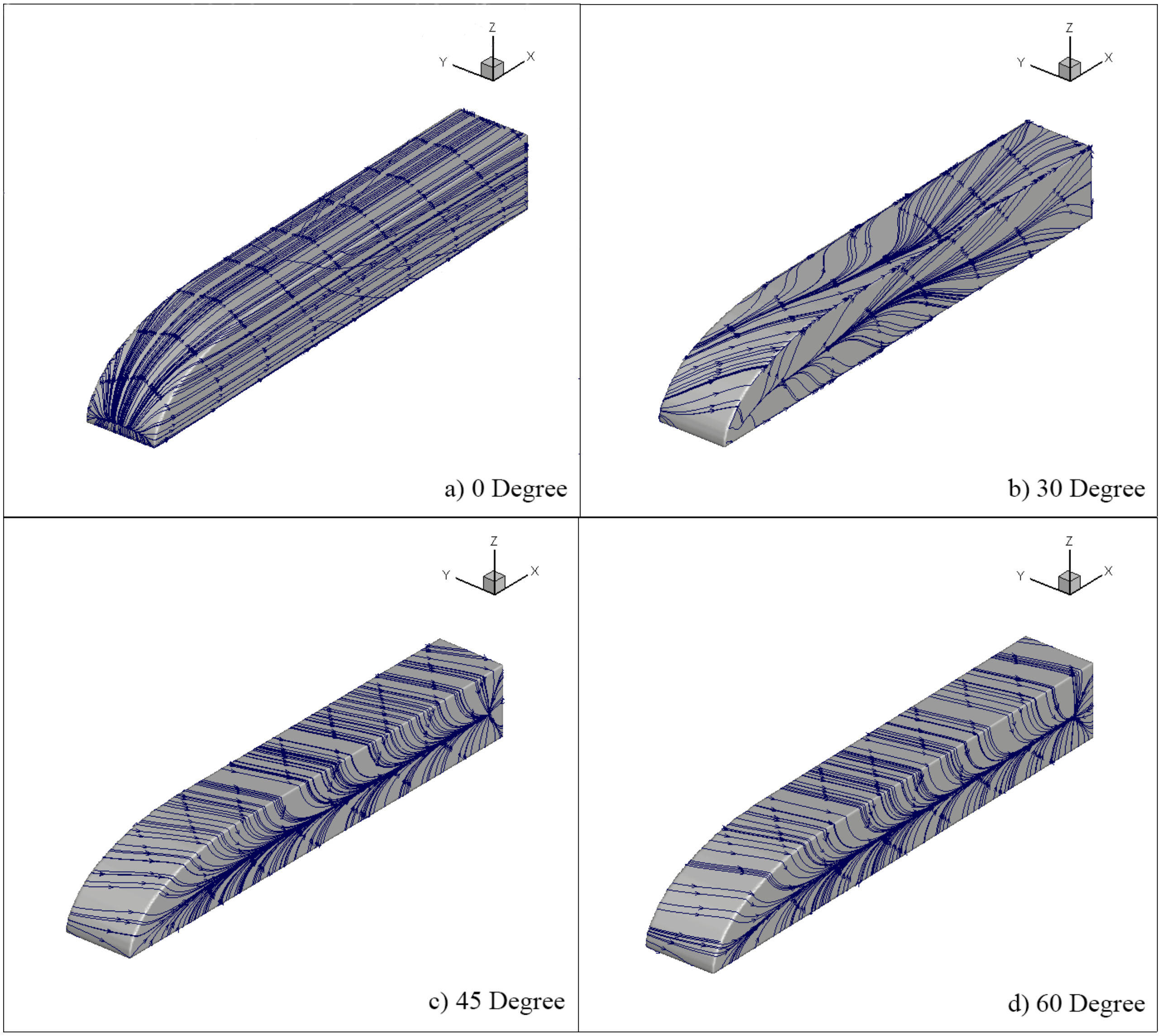

Figure 7. Surface streamlines at the windside wall for different wind direction angles.

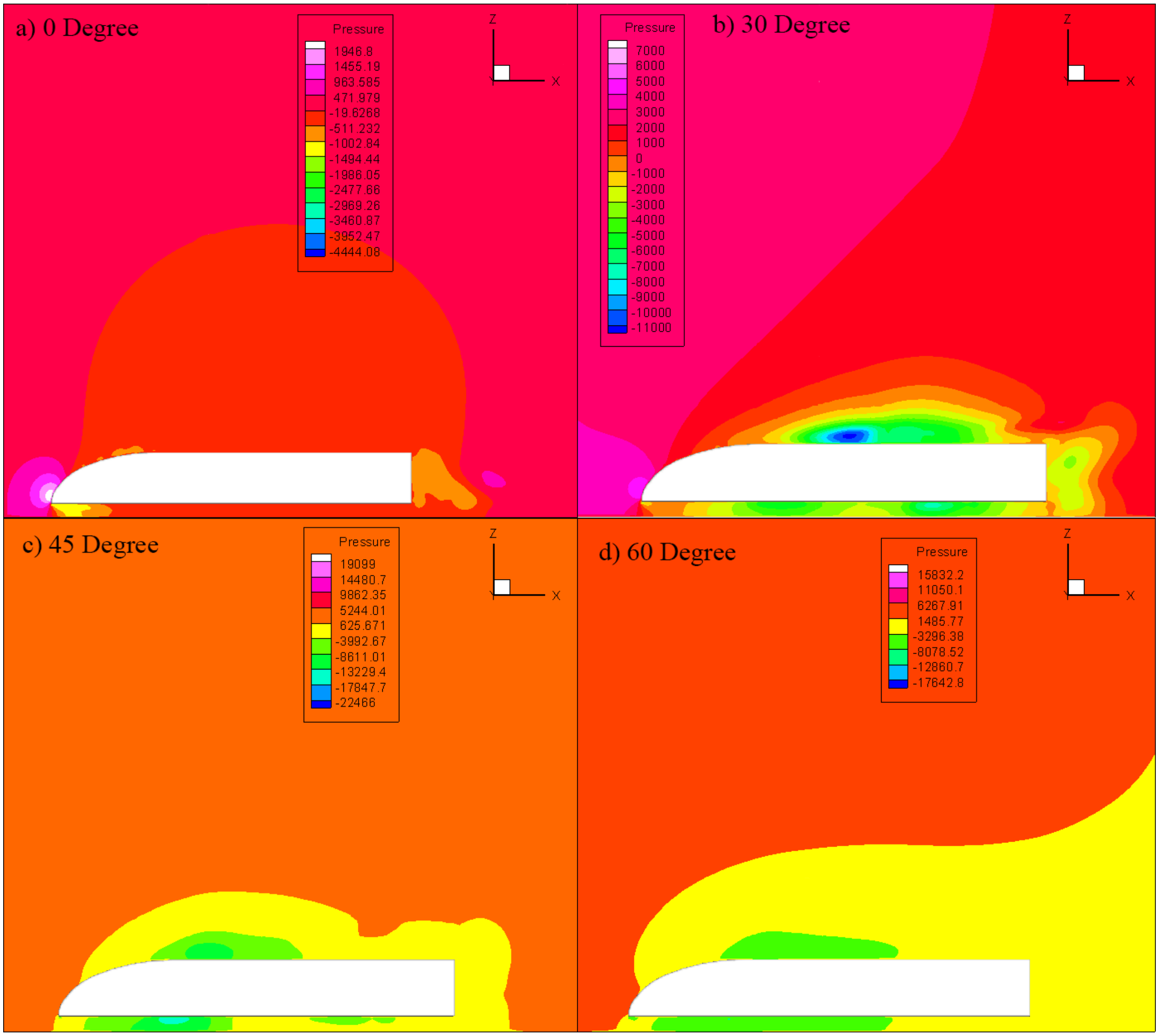

Figure 8. Two dimensional pressure distribution around the train for different wind direction angles.

Moreover, the two dimensional velocity distribution (x-axis velocity: *U*) around the train for different wind direction angles are illustrated in Figure 9. Based on principles, the velocity value close to the train body is lower and at the farther away is higher.

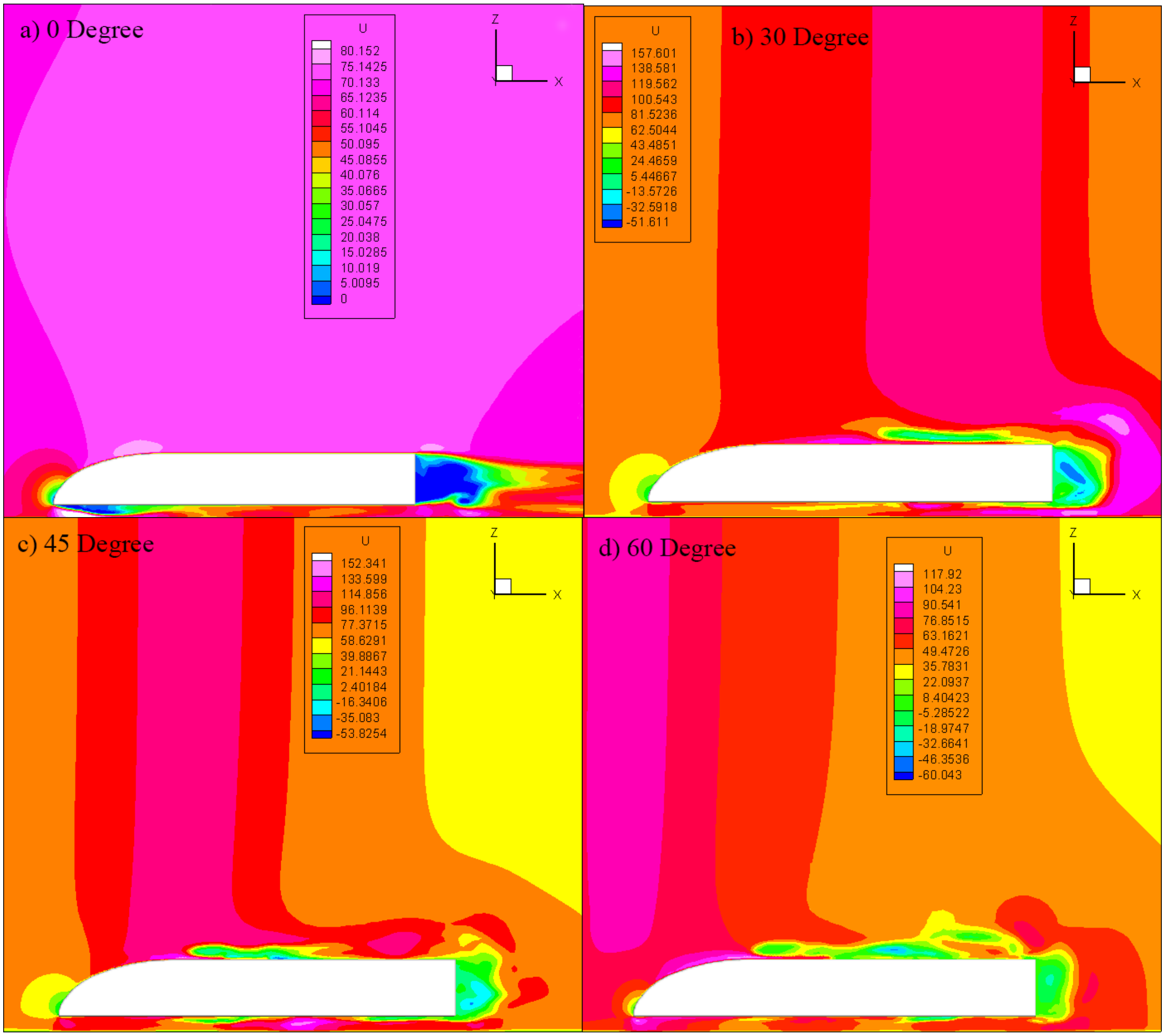

Figure 9. Two dimensional velocity distribution around the train for different wind direction angles.

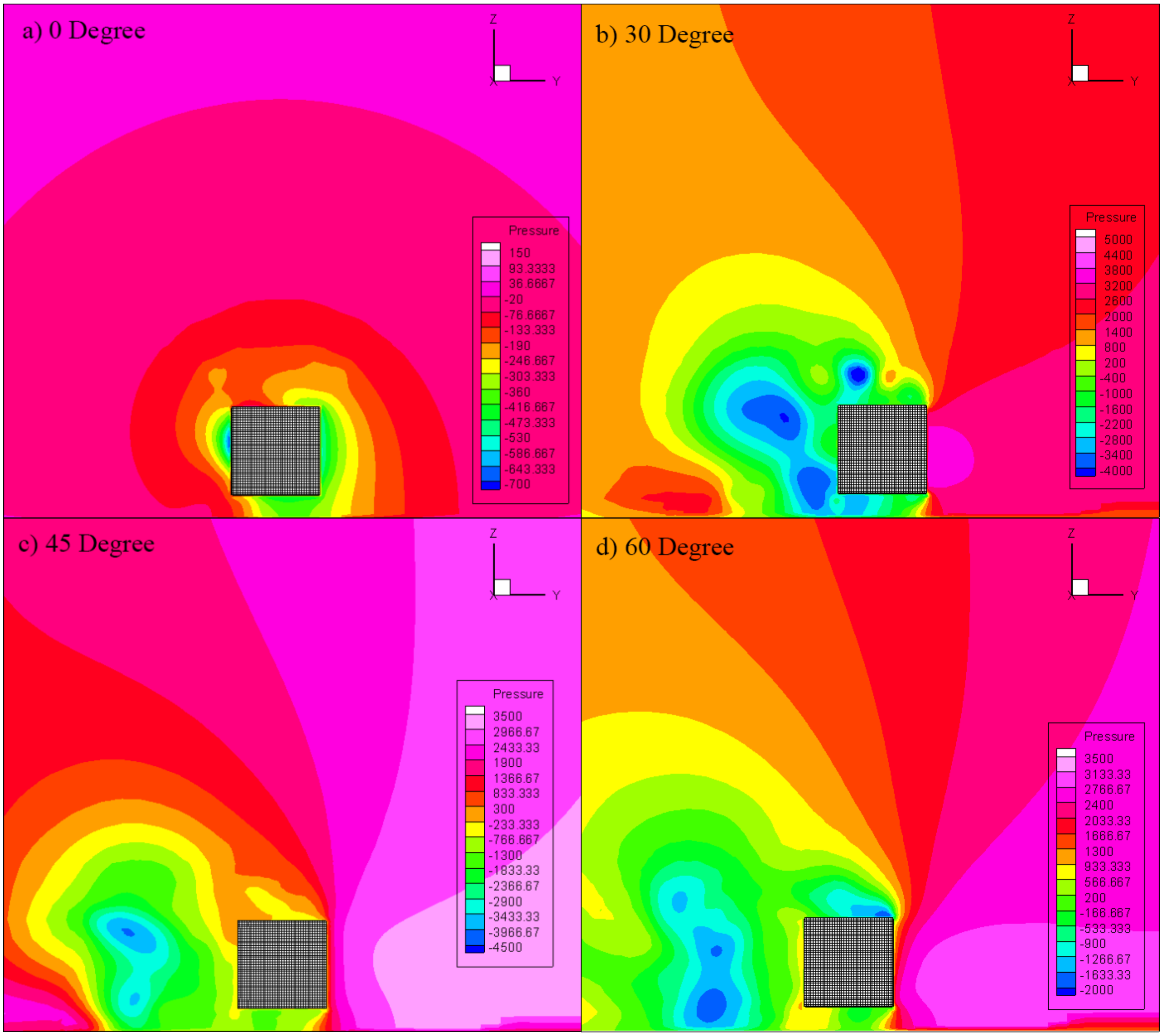

Figure 10. Two dimensional pressure distribution along train cross section for different wind direction angles.

Also, Figure 10 shows the two dimensional pressure distribution along train cross section for different wind direction angles ($\theta = 0^{\circ}$, $30^{\circ}$, $45^{\circ}$ and $60^{\circ}$). All cases illustrate the low region of pressure at the leeside of the train if compared to the windward side.

### 3.2 Turbulent Characteristics

The contours of two different turbulent parameters of turbulent kinetic energy ($k$) and specific dissipation rate ($\omega$) for different wind direction angles are presented in Figures 11 and 12, respectively. As the air stream collides with the generic train model, the regime of air flow stream changes form laminar to turbulent. The zones with turbulent regime and the intensity of turbulence may be identified with kinetic turbulent energy parameter. As shown as Figure 13, in case of $\theta$ = 0°, the highest value of turbulent kinetic energy occurs at the adjacent of the nose of the model. Furthermore, the turbulent kinetic energy reduces level by level as the air flow getting away from the generic train model. It is due to the fact that the air flow speed becomes slower and the disordered flow structure changes to regular flow pattern. In addition, it is clear that the nose of the model has pronounced influence on determining the turbulent region. In the next cases with $\theta$ = 30°, 45° and 60°, the regions with different turbulent kinetic energy are totally irregular as same as the three-dimensional flow structure at this case. It can be observed that the regions near to the train model have higher value of turbulent kinetic energy.

The specific turbulence dissipation is the rate at which turbulence kinetic energy is converted into thermal internal energy per unit volume and time. The values of specific turbulent dissipation rate for the cases $\theta$ = 0°, 30°, 45° and 60° are almost similar with each other. The contours of specific dissipation rate reveal that the maximum value of this parameter occurs at the surface of the nose of the generic high-speed train model.

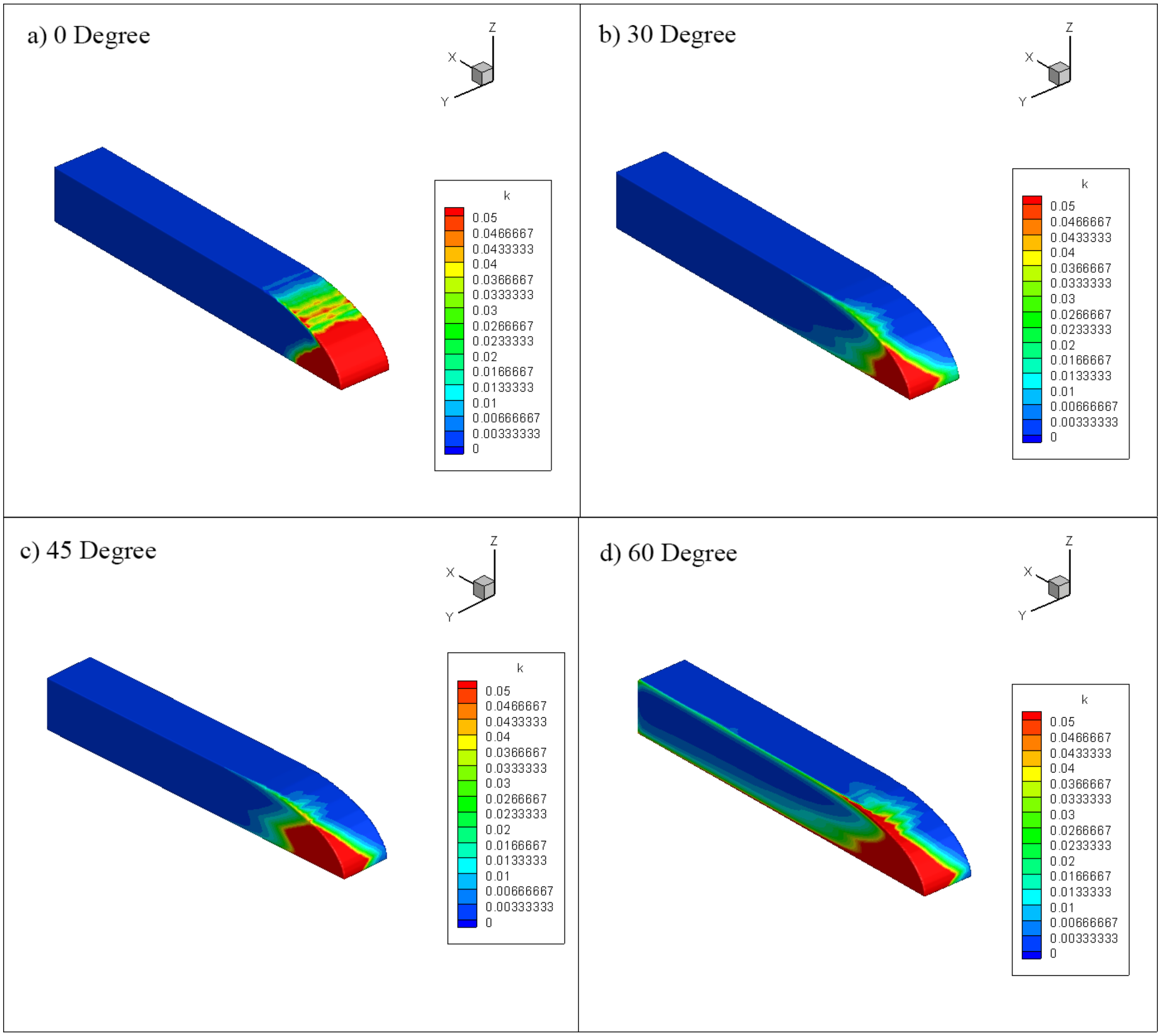

Figure 11. Surface Kinetic Energy for different wind direction angles.

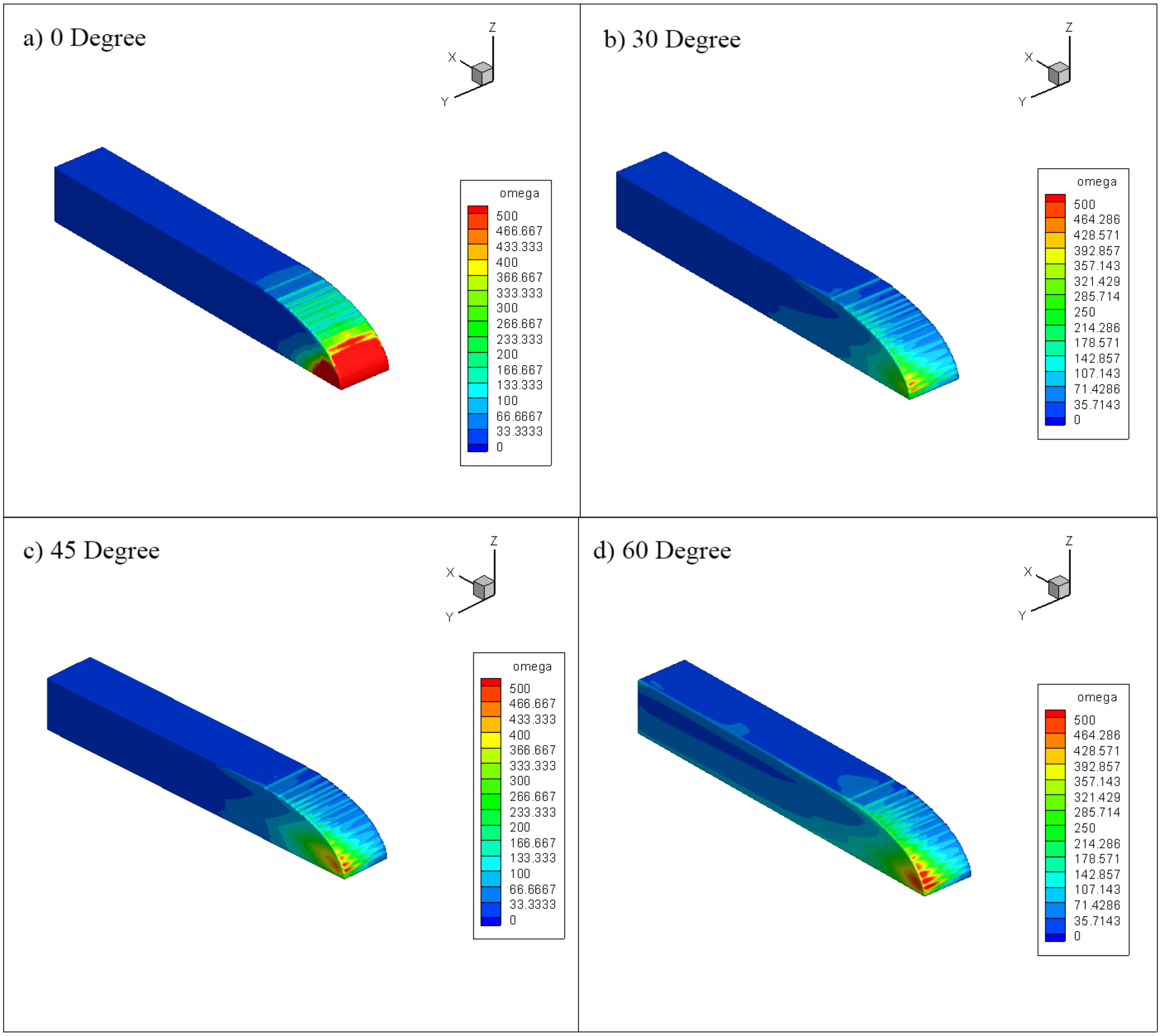

Figure 12. Surface Specific Dissipation Rate for different wind direction angles.

## 3.3. Aerodynamic Forces

The most significant and practical aerodynamic parameters for bluff bodies simulation are lift, drag and side forces. To achieve this, the mentioned forces for train case for different wind directions are defined and estimated, clearly. The lift force and the drag force coefficients are defined as follows (Zhuang, 2015):

$$C_L = \frac{F_L}{\frac{1}{2}\rho U_\infty^2 A_L} \tag{19}$$

and

$$C_D = \frac{F_D}{\frac{1}{2}\rho U_\infty^2 A_D} \tag{20}$$

where, $F_D$ and $F_L$ are the drag and lift forces, $A_D$ and $A_L$ are the surface area of the train in y- and z- directions, respectively. Moreover, the side force coefficient is defined as follows:

$$C_S = \frac{F_S}{\frac{1}{2}\rho U_\infty^2 A_S} \tag{21}$$

where, $F_S$ and $A_S$ are the side force and the side surface area of the train in x- direction, respectively. Also, the pressure coefficient is as follow:

$$C_p = \frac{P - P_\infty}{\frac{1}{2}\rho U_\infty^2} \tag{22}$$

where, $P_\infty$ , $\rho$ and $U_\infty$ are the free stream pressure, density and free stream velocity, respectively.

For verifying the extracted data from this paper, did an approximate comparison between the aerodynamic coefficients obtained from this paper and the (Zhuang, 2015). The geometric conditions of the two papers are as the same and for more accuracy the free stream velocity was considered $U_\infty$ = 20 m/s, as the (Zhuang, 2015). The comparison results is illustrated as Table 1. The comparison show that the a good agreement between the results of this paper and the obtained results from (Zhuang, 2015).

In the following, for different wind directions of this research, i.e. 0˚, 30˚, 45˚ and 60˚ and for 70 m/s free stream velocity, the lift, drag and side aerodynamic coefficients are compared which are illustrated in Table 2. As can be seen from Table 2, when the wind directions increase, the lift, drag and side aerodynamic coefficients increase. Then, the friction and resistance against train movement increase too. Moreover, in Table 3, the time-averaged of the minimum, maximum and average values of the pressure coefficients for mentioned different wind directions are listed and compared. As the same way, with increasing the wind direction of the free stream, the numerical values of the minimum, maximum and the average values of the pressure coefficients increase too. Also, the maximum value of pressure coefficient at 60˚ wind direction has the highest value.

Then, the exact pressure coefficient for different nodes on the train surface as Figure 13 are shown in Figure 14. As shown in Figure 14, the top nodes of the train roof which are the midpoints of the roof to pressure coefficients analysis and compare are considered. The desired values are listed in Figure 14 based on the marked points.

Table 1. Comparison of time-averaged values of the lift and side forces between the paper and the (Zhuang, 2015).

| **Cases** | **This paper** | | **(Zhuang, 2015)** | |
|---|---|---|---|---|
| **Wind directions** | $C_L$ | $C_S$ | $C_L$ | $C_S$ |
| **$\theta$ = 30°** | 0.90 | 0.71 | 0.136 | 0.424 |
| **$\theta$ = 60°** | 1.63 | 1.42 | 0.161 | 1.029 |

Table 2. Time-average aerodynamic force coefficients for different wind directions.

| **Aerodynamic Coeffs.** | **$\overline{C_L}$, Lift Coefficient** | **$\overline{C_D}$, Drag Coefficient** | **$\overline{C_S}$, Side Coefficient** |
|---|---|---|---|
| **0°** | 0.34 | 0.48 | 0.01 |
| **30°** | 0.90 | 0.71 | 2.25 |
| **45°** | 1.06 | 0.92 | 2.61 |
| **60°** | 1.63 | 1.42 | 3.02 |

Table 3. Time-average minimum, maximum and average pressure coefficients for different wind directions.

| **Pressure Coeffs.** | $\overline{C_{P,min}}$ | $\overline{C_{P,max}}$ | $\overline{C_{P,ave}}$ |
|---|---|---|---|
| **0°** | -1.91 | 0.98 | -0.10 |
| **30°** | -4.10 | 2.88 | -0.03 |
| **45°** | -6.72 | 4.56 | -0.40 |
| **60°** | -7.80 | 4.66 | -0.33 |

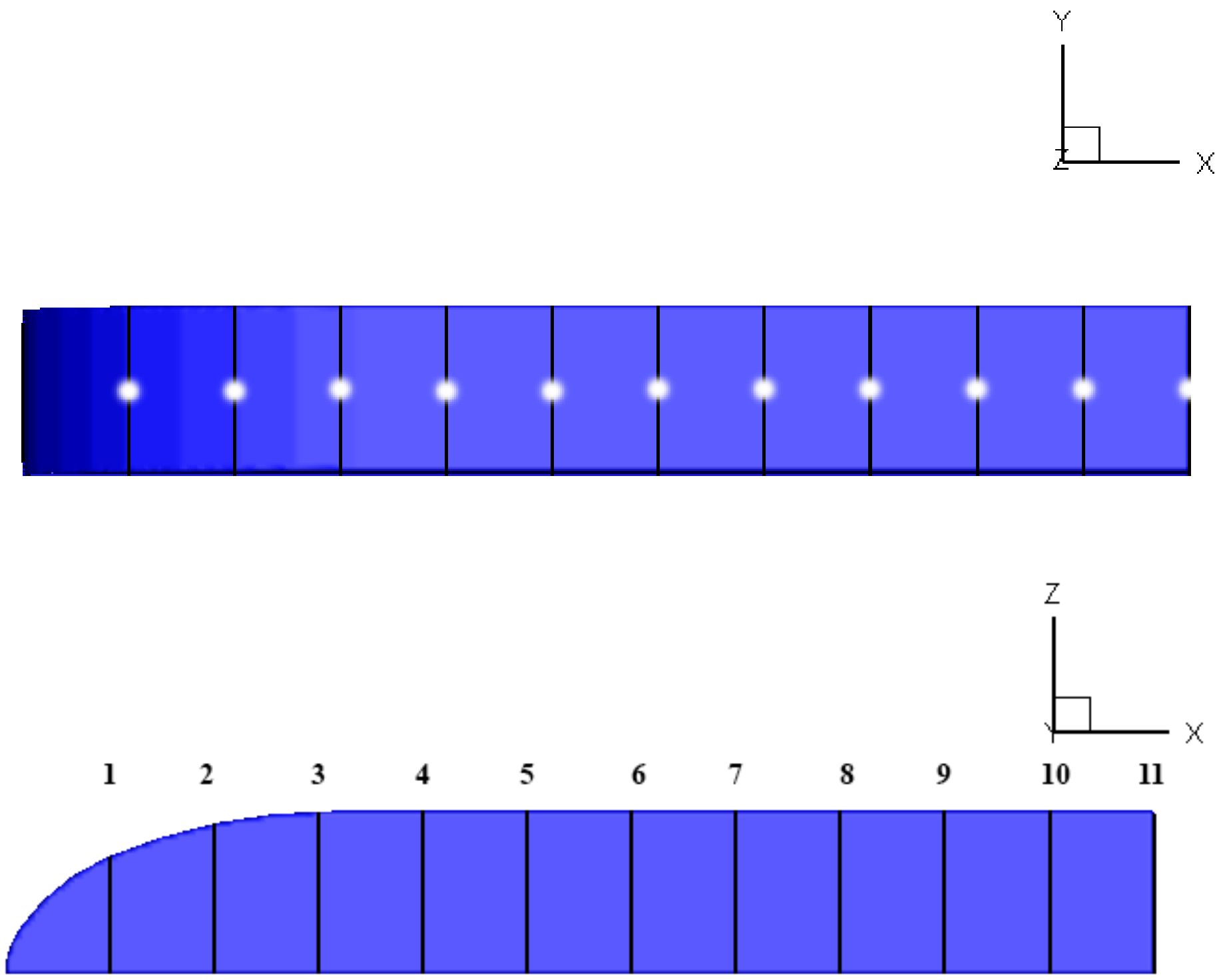

Figure 13. The nodes of train surface for pressure coefficient in Figure 14.

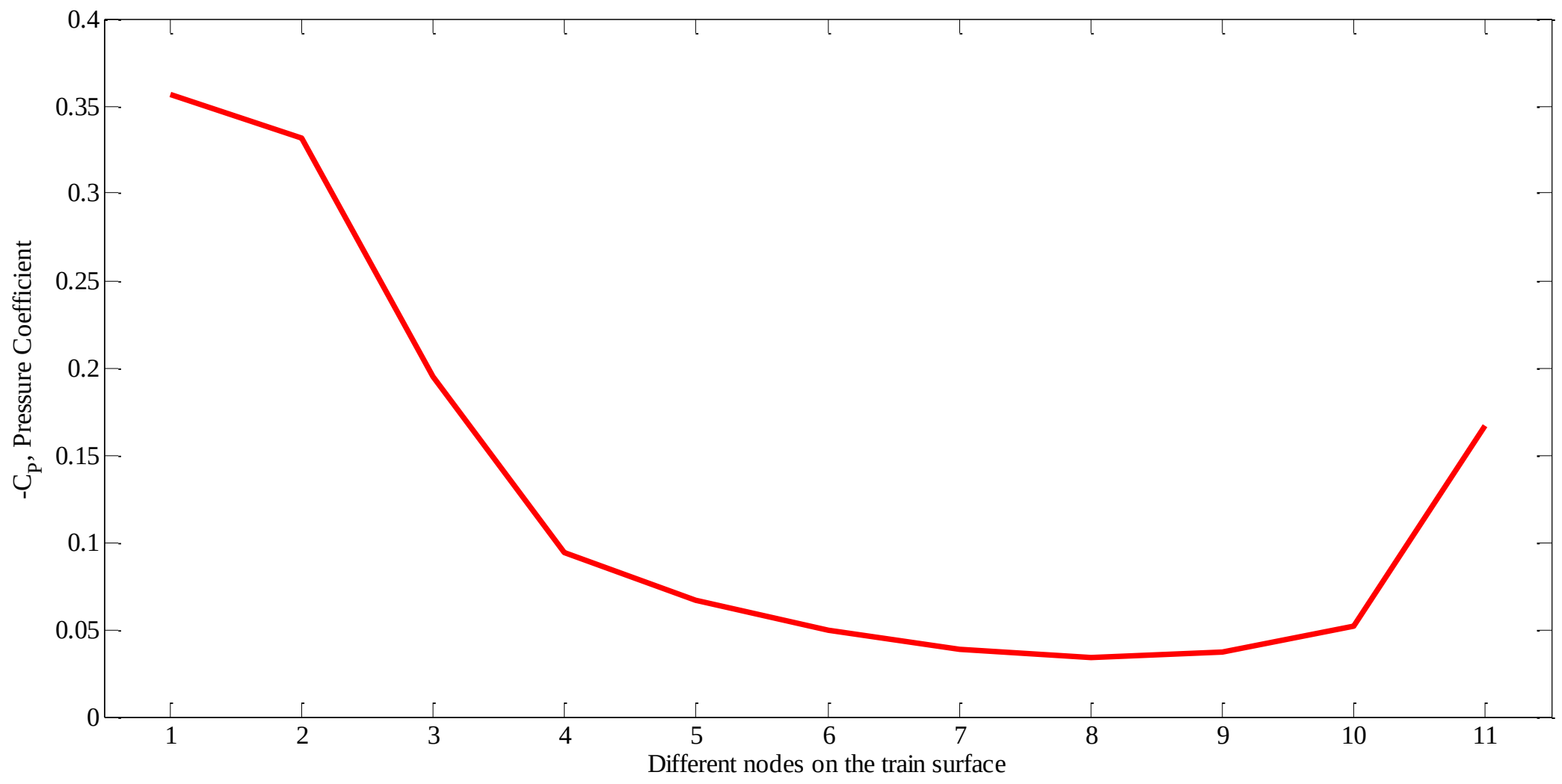

Figure 14. Pressure coefficients for different nodes on the train surface.

In this part, two comparisons on some aerodynamic key parameters for five different free stream velocity as 50, 60, 70, 80 and 90 m/s (and consequently five different Reynolds numbers as $1.9\times10^6$, $2.2\times10^6$, $2.6\times10^6$, $3.0\times10^6$ and $3.4\times10^6$) are done.

In the following, for the different free stream velocity, the lift, drag and side aerodynamic coefficients are compared which are illustrated in Table 4. As it can be seen from Table 4, when the wind velocity increases, the lift, drag and side aerodynamic coefficients increase. Then, the friction and resistance against train movement increase too.

Moreover, in Table 5, the time-averaged of the minimum, maximum and average values of the pressure coefficients for the different wind velocity and Reynolds numbers are listed and compared. According to the results, the maximum value of the pressure coefficient is related to the 90 m/s velocity.

In the following, the aerodynamic drag coefficient for some points during the train length is illustrated in Figure 15 based on the point which shown in Figure 13. As can be seen, the aerodynamic drag for 15 points during the train length for 5 air flow velocity (50, 60, 70, 80 and 90 m/s) is analyzed. With increasing the air flow velocity, the aerodynamic drag for similar points increase too. Also, from the nose to the end of the train, the aerodynamic drag has a downward trend. Moreover, The maximum value of the drag coefficient is related to the case of 90 m/s and occurs at the nose of the train.

Table 4. Time-average aerodynamic force coefficients for different free stream velocity.

| **Aerodynamic Coeffs.** | $\overline{C_L}$, **Lift Coefficient** | $\overline{C_D}$, **Drag Coefficient** | $\overline{C_S}$, **Side Coefficient** |
|---|---|---|---|
| **50 (m/s)** | 0.23 | 0.36 | 0.012 |
| **60 (m/s)** | 0.29 | 0.41 | 0.019 |
| **70 (m/s)** | 0.36 | 0.48 | 0.018 |
| **80 (m/s)** | 0.46 | 0.53 | 0.024 |
| **90 (m/s)** | 0.53 | 0.69 | 0.027 |

Table 5. Time-average minimum, maximum and average pressure coefficient for different free stream velocity.

| **Pressure Coeffs.** | $\overline{C_{P,min}}$ | $\overline{C_{P,max}}$ | $\overline{C_{P,ave}}$ |
|---|---|---|---|
| **50 (m/s)** | -0.97 | 0.50 | -0.05 |
| **60 (m/s)** | -1.40 | 0.72 | -0.06 |
| **70 (m/s)** | -1.91 | 0.98 | -0.10 |
| **80 (m/s)** | -2.50 | 1.29 | -0.13 |
| **90 (m/s)** | -3.16 | 1.63 | -0.17 |

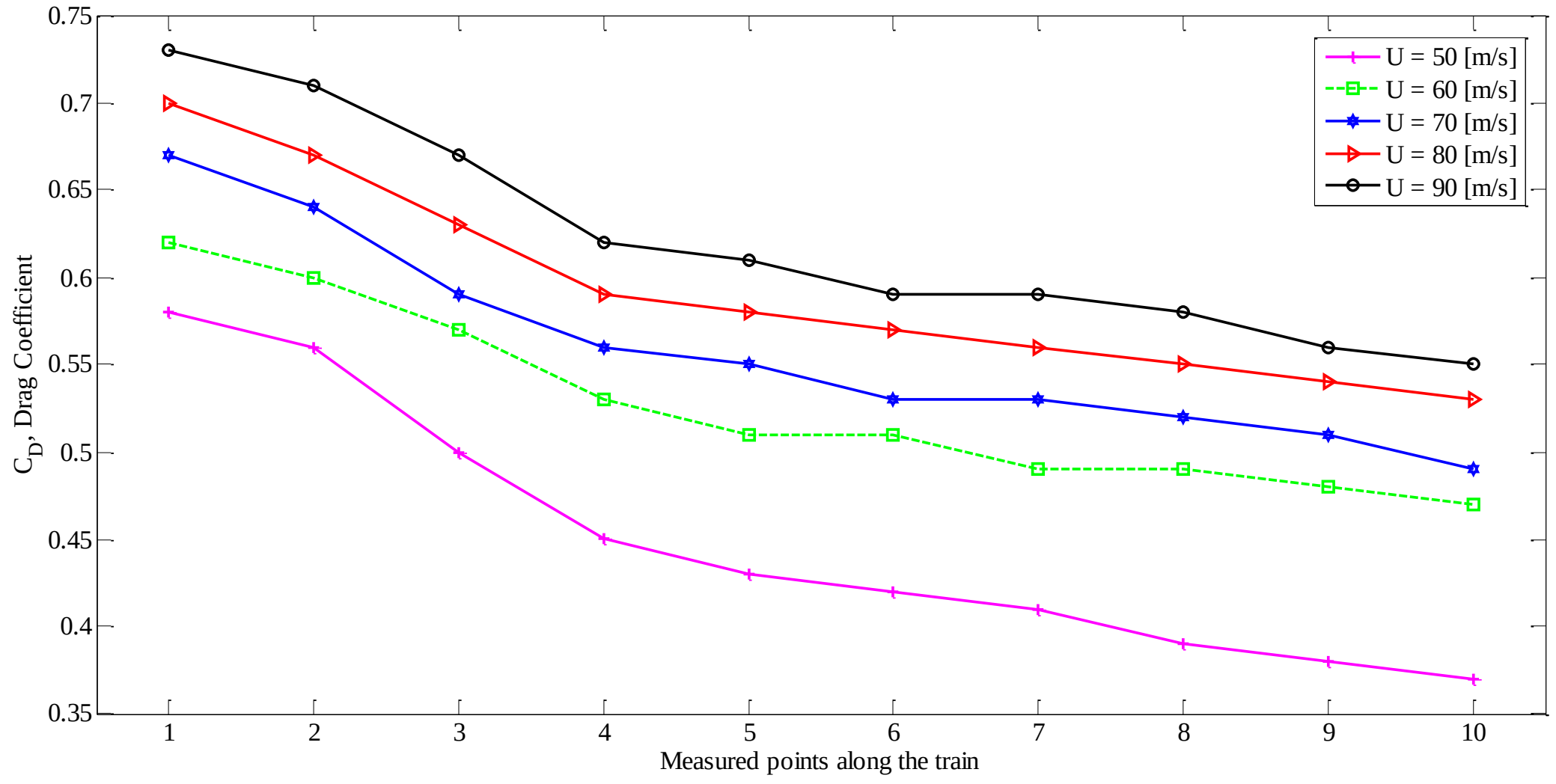

Figure 15. Aerodynamic drag coefficients during the train length.

# 4. Prediction Methods

## 4.1. Introduction

In this section using Gene Expression Programming (GEP), Gaussian Process Regression (GPR) and random forest (RF) methods, the aerodynamic parameters as drag, lift and side forces and also minimum and maximum values of pressure coefficients are predicted for mentioned wind directions (for $\theta = 0^\circ$ to $60^\circ$) and velocity (for 50 m/s to 90 m/s). The statistical parameters of utilized data for both wind direction and free velocity coefficients are presented at Table 6.

Table 6. Statistical characteristics of the utilized data.

| | Variable | mean | minimum | maximum | standard deviation | coefficient of variation | skewness | Correlation |
|---|---|---|---|---|---|---|---|---|
| | Wind direction | 30.00 | 0.00 | 60.00 | 17.75 | 0.59 | 0.00 | 1 |
| | $C_L$ | 0.89 | 0.34 | 1.63 | 0.34 | 0.38 | 0.44 | 0.978 |
| | $C_D$ | 0.83 | 0.48 | 1.42 | 0.27 | 0.33 | 0.95 | 0.933 |
| **Wind direction** | $C_S$ | 1.88 | 0.02 | 3.02 | 0.91 | 0.48 | -0.55 | 0.975 |
| | $C_{P,min}$ | -4.69 | -7.81 | -1.91 | 1.98 | -0.42 | -0.20 | -0.986 |
| | $C_{P,max}$ | 3.06 | 0.98 | 4.67 | 1.26 | 0.41 | -0.03 | 0.987 |
| | Free stream | 70.00 | 50.00 | 90.00 | 11.98 | 0.17 | 0.00 | 1 |
| | $C_L$ | 0.37 | 0.23 | 0.53 | 0.10 | 0.26 | 0.20 | 0.994 |
| | $C_D$ | 0.49 | 0.36 | 0.69 | 0.09 | 0.19 | 0.59 | 0.975 |
| **Free stream Velocity** | $C_S$ | 0.02 | 0.01 | 0.03 | 0.00 | 0.25 | 0.08 | 0.928 |
| | $C_{P,min}$ | -1.97 | -3.17 | -0.97 | 0.68 | -0.35 | -0.05 | -0.998 |
| | $C_{P,max}$ | 1.02 | 0.50 | 1.63 | 0.33 | 0.33 | 0.22 | 0.995 |

As can be seen clearly from Table 6, $C_D$ has the greatest skewness in both wind direction and free stream velocity cases. Moreover, $C_L$ indicates skewed distribution.

### 4.2. Models Performance evaluation parameters

Predictive performances of mentioned models were presented as Correlation coefficient (CC), Root mean squared error (RMSE) and Relative absolute error (RAE). These statistics are presented as follows (S. Samadianfard, Majnooni-Heris, A., Qasem, S. N., Kisi, O., Shamshirband, S. & Chau, K. W., 2019) and (Qasem, 2019):

I: Correlation coefficient (CC), expressed as:

$$CC = \frac{\left(\sum_{i=1}^{n} O_i P_i - \frac{1}{n}\sum_{i=1}^{n} O_i \sum_{i=1}^{n} P_i\right)}{\left(\sum_{i=1}^{n} O_i^2 - \frac{1}{n}\left(\sum_{i=1}^{n} O_i\right)^2\right)\left(\sum_{i=1}^{n} P_i^2 - \frac{1}{n}\left(\sum_{i=1}^{n} P_i\right)^2\right)} \tag{1}$$

II: Root Mean Squared Error (RMSE) follows as:

$$RMSE = \sqrt{\frac{1}{n}\sum_{i=1}^{n}\left(P_i - O_i\right)^2} \tag{2}$$

III: Relative Absolute Error (RAE) stated as:

$$RAE = \frac{\sum_{i=1}^{n}\left|P_i - O_i\right|}{\sum_{i=1}^{n}\left|O_i - \overline{O}\right|} \tag{3}$$

Where, $O_i$ and $P_i$ are the observed and predicted $i^{th}$ value.

### 4.3. Prediction Results and Discussion

In the current research, the coefficients of $C_D$, $C_L$, $C_S$, $C_{P,min}$ and $C_{P,max}$ were estimated in both wind direction and free stream velocity cases using Gene Expression Programming (GEP), Gaussian Process Regression (GPR) and random forest (RF) methods. It should be noted that there is no straightforward way of splitting training and testing data. For example, the study of (Kurup, 2014) used a total of 63% of their data for model development, whereas (S. Samadianfard, Delirhasannia, R., Kisi, O., & Agirre-Basurko, E., 2013) and (S. Samadianfard, Sattari, M. T., Kisi, O., & Kazemi, H., 2014) used 67% of total data, and (Deo, 2018) used 70% of total data to develop their models. Thus, to develop the studied GEP, GPR and RF models for estimation of aerodynamic coefficients, we divided the data into training (70%) and testing (30%). Additionally, the parameters of GEP are displayed in Table 7. Hence, the obtained results of the statistical parameters for GEP, GPR and RF models in the test phase are given in Table 8.

Table 7. Parameters of the GEP model.

| Parameter | Value |
|---|---|
| Chromosomes | 30 |
| Head size | 7 |
| Number of Genes | 3 |
| Linking Function | Addition (+) |
| Mutation Rate | 0.044 |
| Inversion Rate | 0.1 |
| One-Point Recombination Rate | 0.3 |
| Two-Point Recombination Rate | 0.3 |
| Gene Recombination Rate | 0.1 |
| Gene Transposition Rate | 0.1 |
| Used functions | +, -, ×, ÷, power, Ln, sin, cosine, |

Table 8. General results of the computations for the studied models.

| | **coefficients** | **GEP** | | | **GPR** | | |
|---|---|---|---|---|---|---|---|
| | | **CC** | **RMSE** | **RAE** | **CC** | **RMSE** | **RAE** |
| **Wind direction** | $C_L$ | 0.9878 | 0.0533 | 0.1296 | 0.9796 | 0.0517 | 0.1902 |
| | $C_D$ | 0.9937 | 0.0325 | 0.1013 | 0.9846 | 0.0653 | 0.1815 |
| | $C_S$ | 0.9947 | 0.0884 | 0.1060 | 0.9968 | 0.0896 | 0.0973 |
| | $C_{P,min}$ | 0.9984 | 0.1134 | 0.0509 | 0.9966 | 0.2381 | 0.1094 |
| | $C_{P,max}$ | 0.9991 | 0.0565 | 0.0426 | 0.9979 | 0.1267 | 0.0912 |
| **Free stream Velocity** | $C_L$ | 0.9986 | 0.0048 | 0.0584 | 0.9941 | 0.0118 | 0.1277 |
| | $C_D$ | 0.9901 | 0.0126 | 0.1684 | 0.9887 | 0.0174 | 0.1755 |
| | $C_S$ | 0.8999 | 0.0019 | 0.4135 | 0.9042 | 0.0018 | 0.3835 |

| | | | | | | |
|---|---|---|---|---|---|---|
| $C_{P,min}$ | 0.9976 | 0.0668 | 0.0669 | 0.9941 | 0.0937 | 0.1286 |
| $C_{P,max}$ | 0.9977 | 0.0196 | 0.0656 | 0.9914 | 0.0478 | 0.1008 |

As it can be seen from Table 8, RF had the best performance in estimation of all aerodynamic coefficients in the case of wind direction. In other words, RF in the case of wind direction with CC values of 0.9982, 0.9988, 0.9990, 0.9994, 0.9993, RMSE values of 0.0235, 0.0169, 0.0469, 0.0733, 0.0503 and RAE values of 0.0635, 0.0493, 0.0457, 0.0336, 0.0347 presented more accurate estimations of $C_L$, $C_D$, $C_S$, $C_{P,min}$, $C_{P,max}$, respectively comparing to GEP and GPR models. So, it can be selected as the best among studied models for estimating aerodynamic coefficients in the case of wind direction. Furthermore, somehow different trend was seen in the case of free stream velocity. In this case, GEP estimated $C_L$, $C_{P,max}$ more accurately than GPR and RF models. In other words, GEP with CC values of 0.9986, 0.9977, RMSE values of 0.0048, 0.0196 and RAE values of 0.0584, 0.0656 proved itself as the most precise and powerful model for $C_L$, $C_{P,max}$ estimation and had more better performance in comparison to GPR and RF models. But in estimating $C_D$, $C_S$, $C_{P,min}$ coefficients, RF with CC values of 0.9984, 0.9227, 0.9967, RMSE values of 0.0058, 0.0017, 0.0527 and RAE values of 0.0781, 0.3002, 0.0903 was selected as superior model as it was chosen the best in estimation of aerodynamic coefficients in the case of wind direction.

Moreover, the statistical parameters of SVR, SVR-FOA, GEP and RF models are presented as bar chart in Figure 16. It is clear from this figure that SVR-FOA has higher capability in accurate estimation of SBAHC. Moreover, Figures 16 to 21 show the estimation results of the GEP, GPR and FR models in both wind direction and free stream velocity cases. It can be comprehended from these figures that the estimates of RF are in better agreement than other considered models. Furthermore, GEP may be selected as the second best.

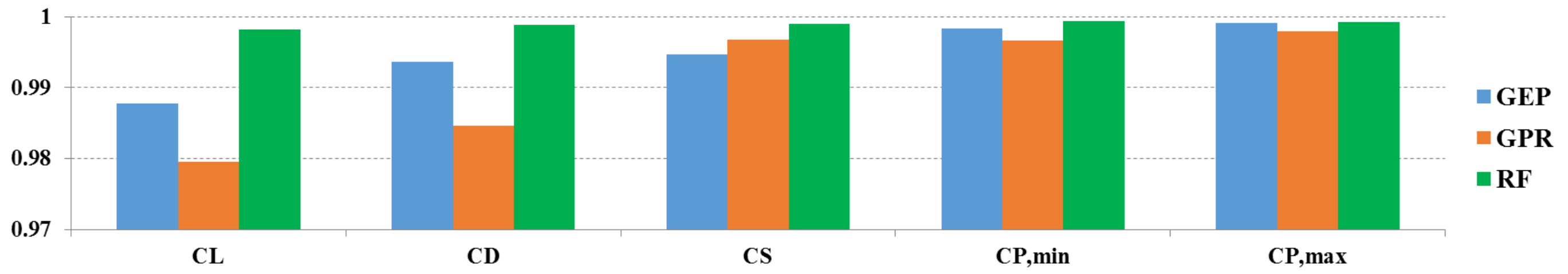

Figure 16. Bar graphs of the CC values (Wind direction).

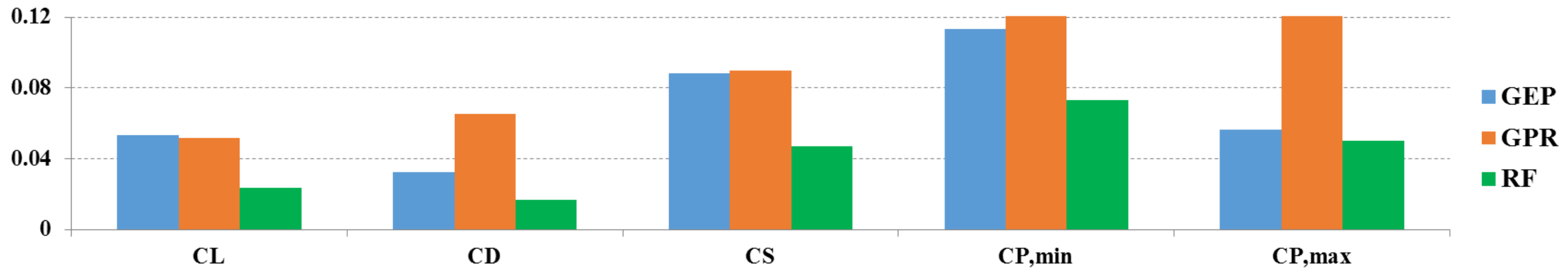

Figure 17. Bar graphs of the RMSE values (Wind direction).

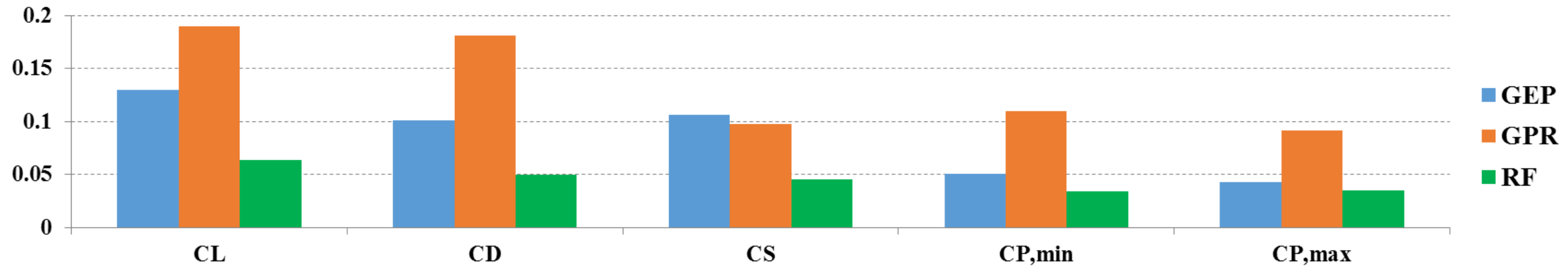

Figure 18. Bar graphs of the RAE values (Wind direction).

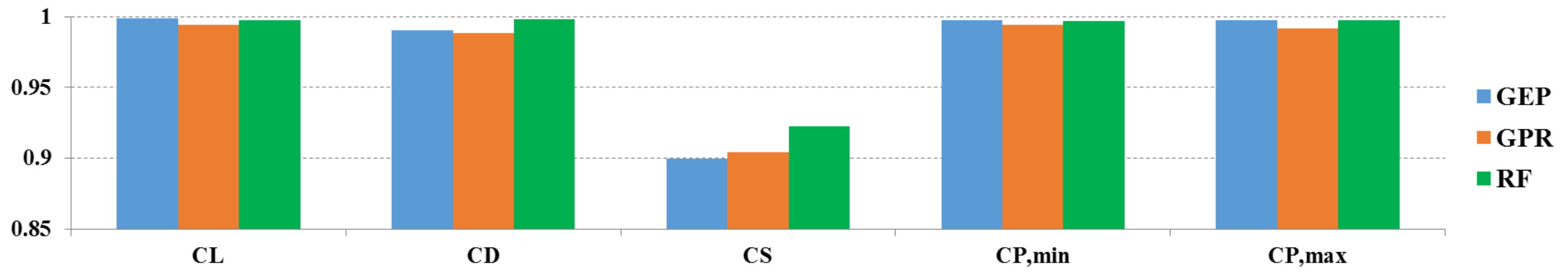

Figure 19. Bar graphs of the CC values (Free stream Velocity).

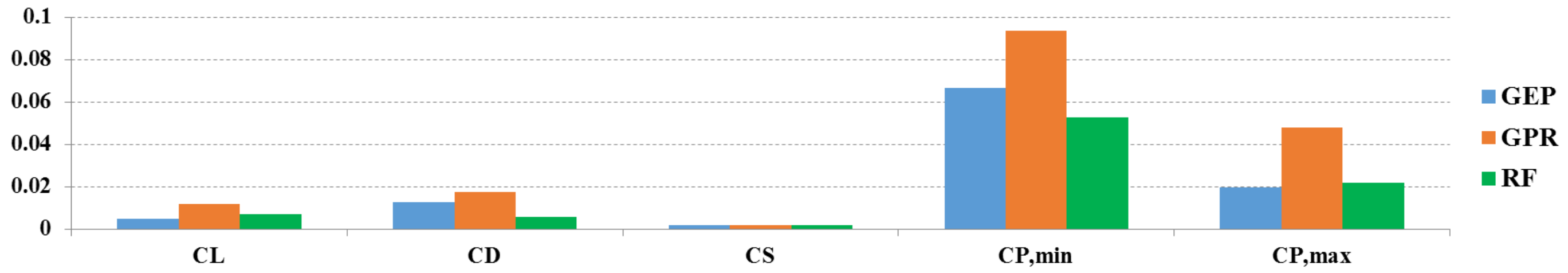

Figure 20. Bar graphs of the RMSE values (Free stream Velocity).

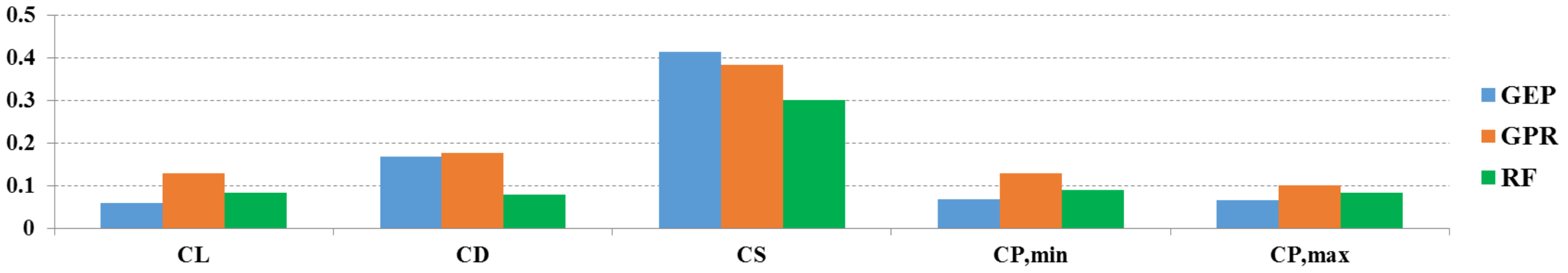

Figure 21. Bar graphs of the RAE values (Free stream Velocity).

Also, the variations of estimated coefficients are illustrated at Figures 22 and 23. Additionally, Figures 24 and 25 indicate scatter plots of estimated coefficients versus observed ones in both wind direction and free stream velocity cases with GEP, GPR and RF models. It is obvious that due to less scattered points, the estimated values of RF are more accurate than GEP and GPR in most cases.

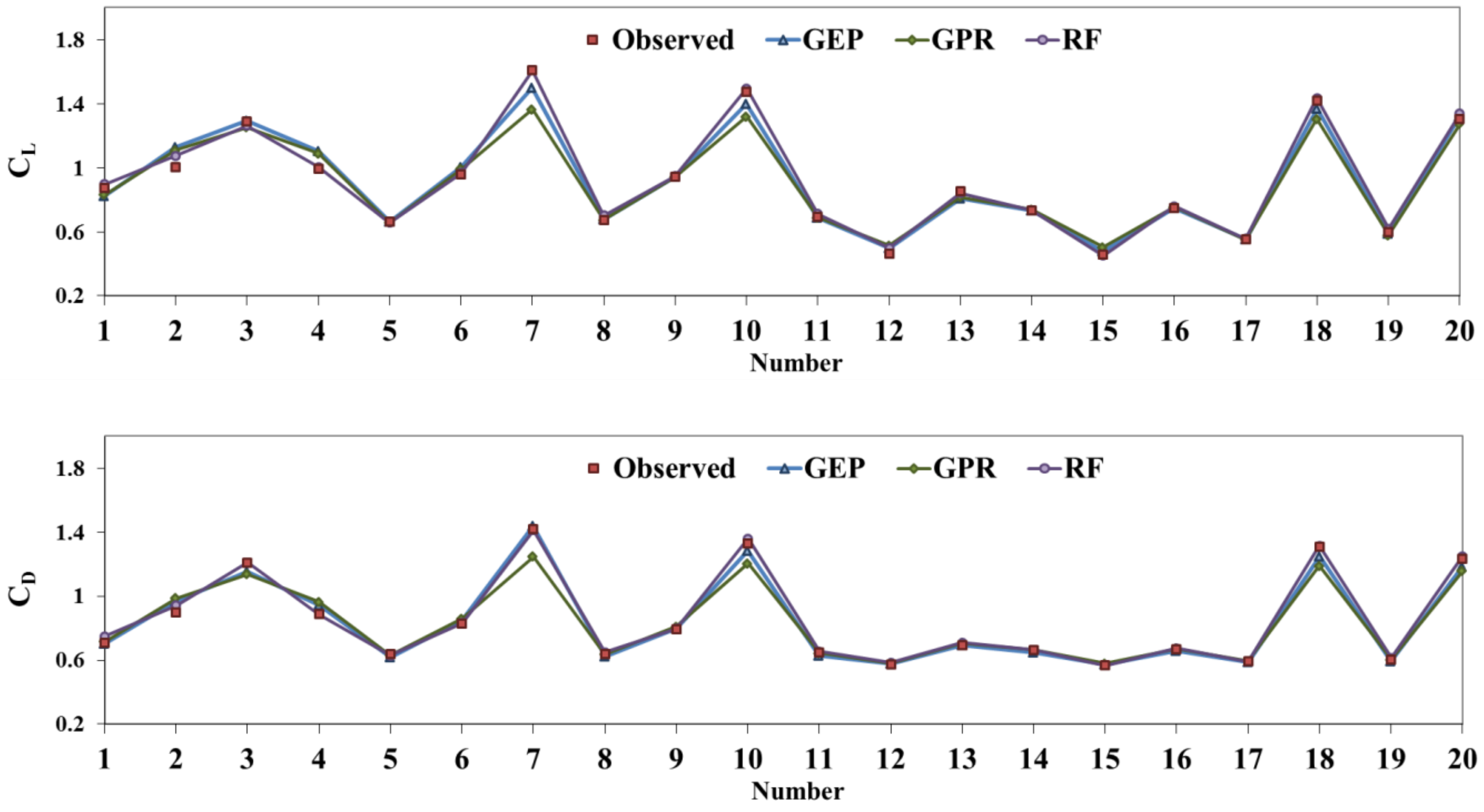

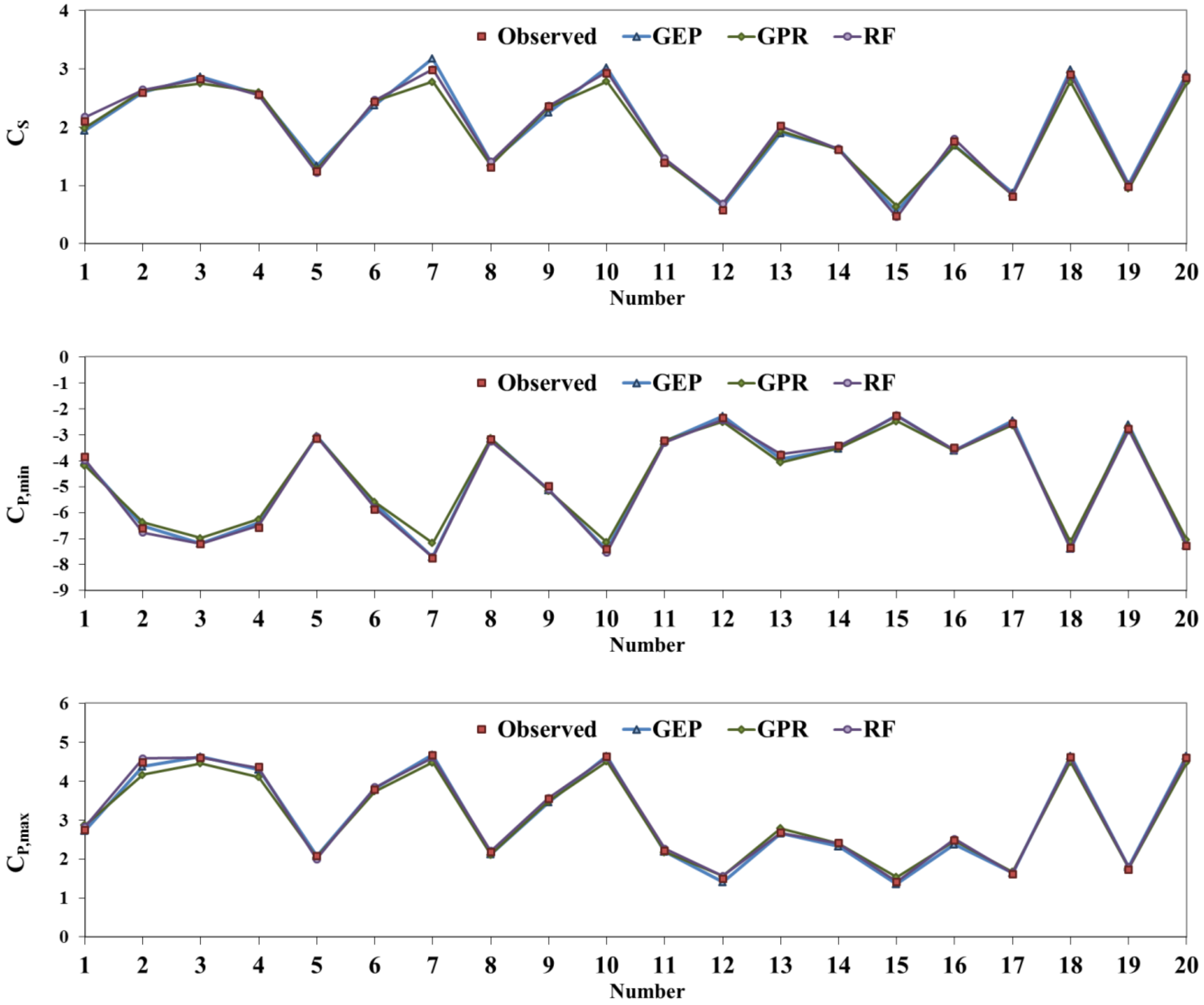

Figure 22. Observed and estimated coefficients (Wind direction).

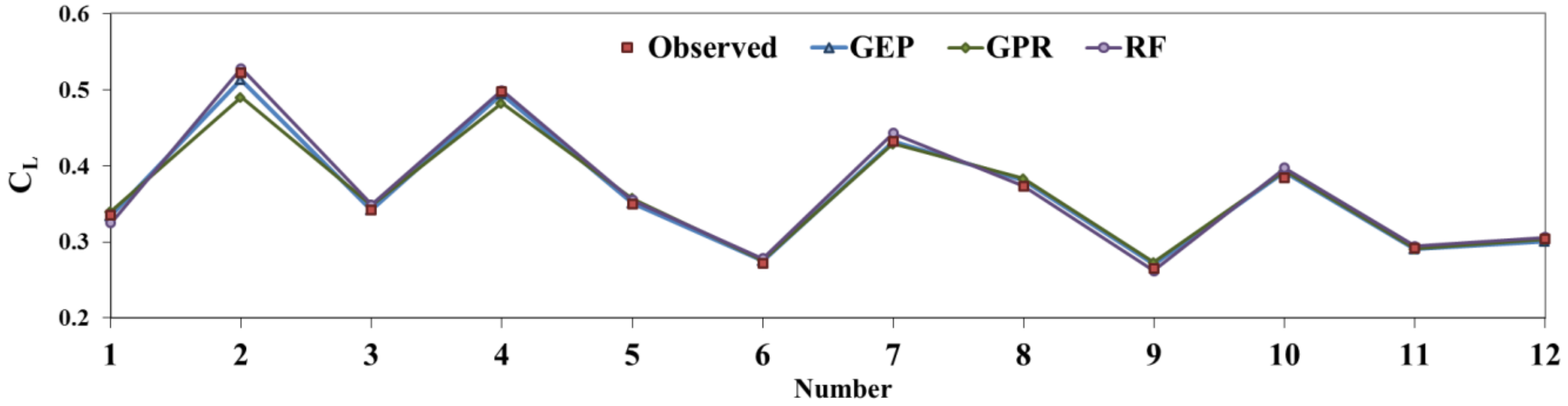

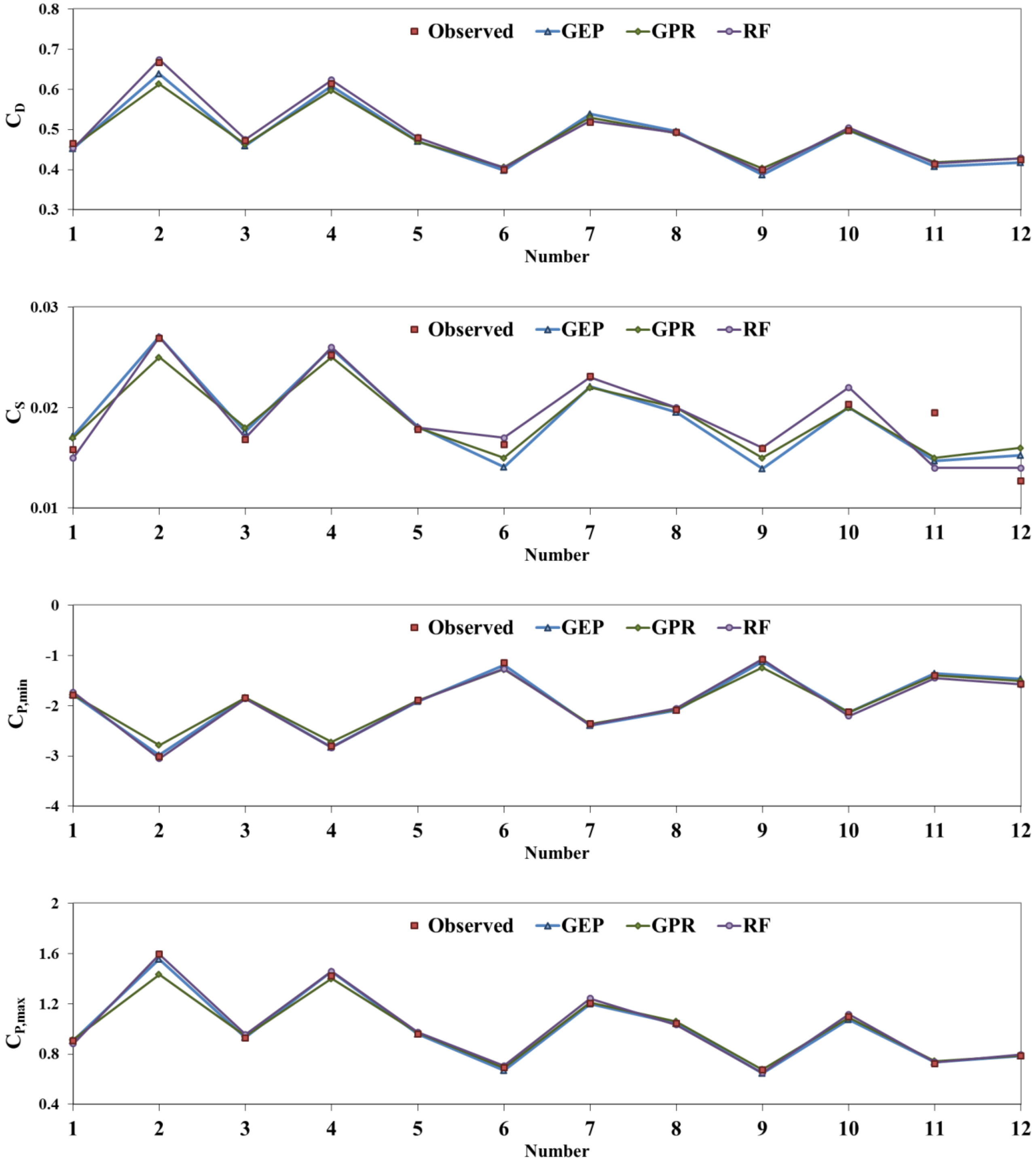

Figure 23. Observed and estimated coefficients (Free stream Velocity).

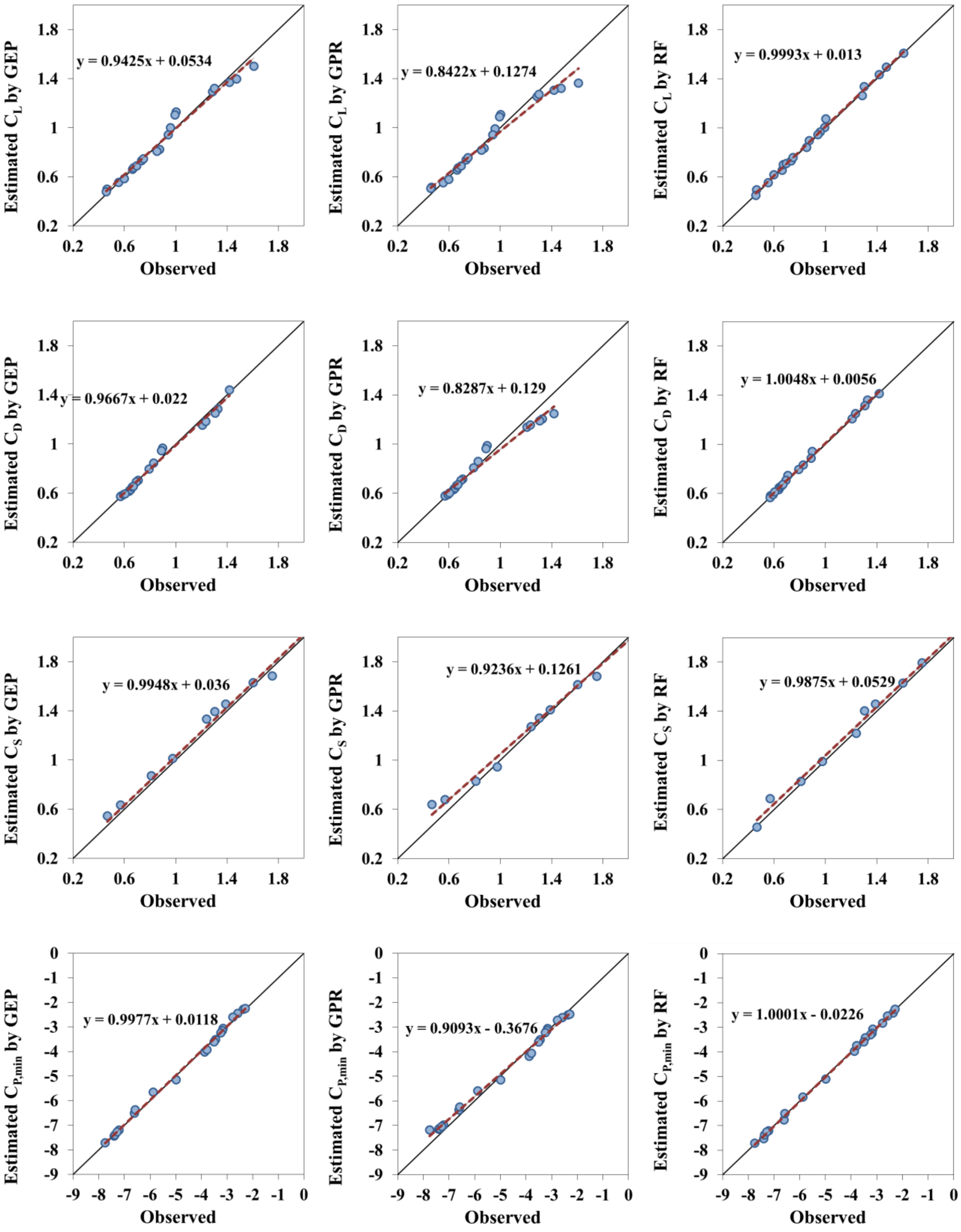
y = 0.9425x + 0.0534
Estimated C_L by GEP
Observed
y = 0.8422x + 0.1274
Estimated C_L by GPR
Observed
y = 0.9993x + 0.013
Estimated C_L by RF
Observed
y = 0.9667x + 0.022
Estimated C_D by GEP
Observed
y = 0.8287x + 0.129
Estimated C_D by GPR
Observed
y = 1.0048x + 0.0056
Estimated C_D by RF
Observed
y = 0.9948x + 0.036
Estimated C_S by GEP
Observed
y = 0.9236x + 0.1261
Estimated C_S by GPR
Observed
y = 0.9875x + 0.0529
Estimated C_S by RF
Observed
y = 0.9977x + 0.0118
Estimated C_P,min by GEP
Observed
y = 0.9093x - 0.3676
Estimated C_P,min by GPR
Observed
y = 1.0001x - 0.0226
Estimated C_P,min by RF
Observed

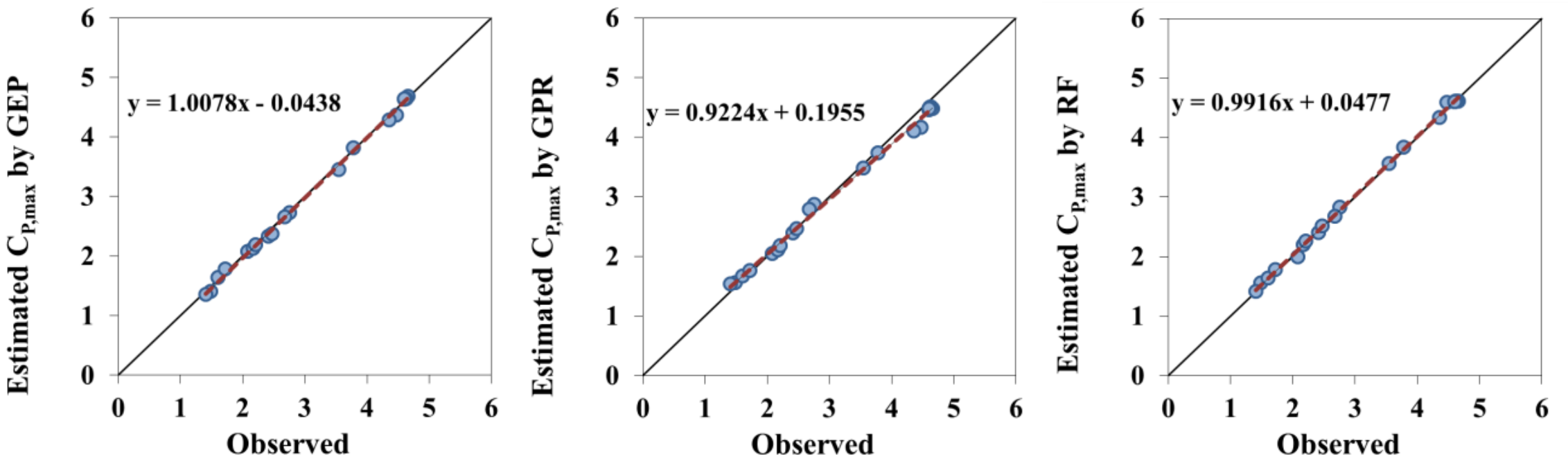

Figure 24. The scatter plots of observed and estimated coefficients (Wind direction).

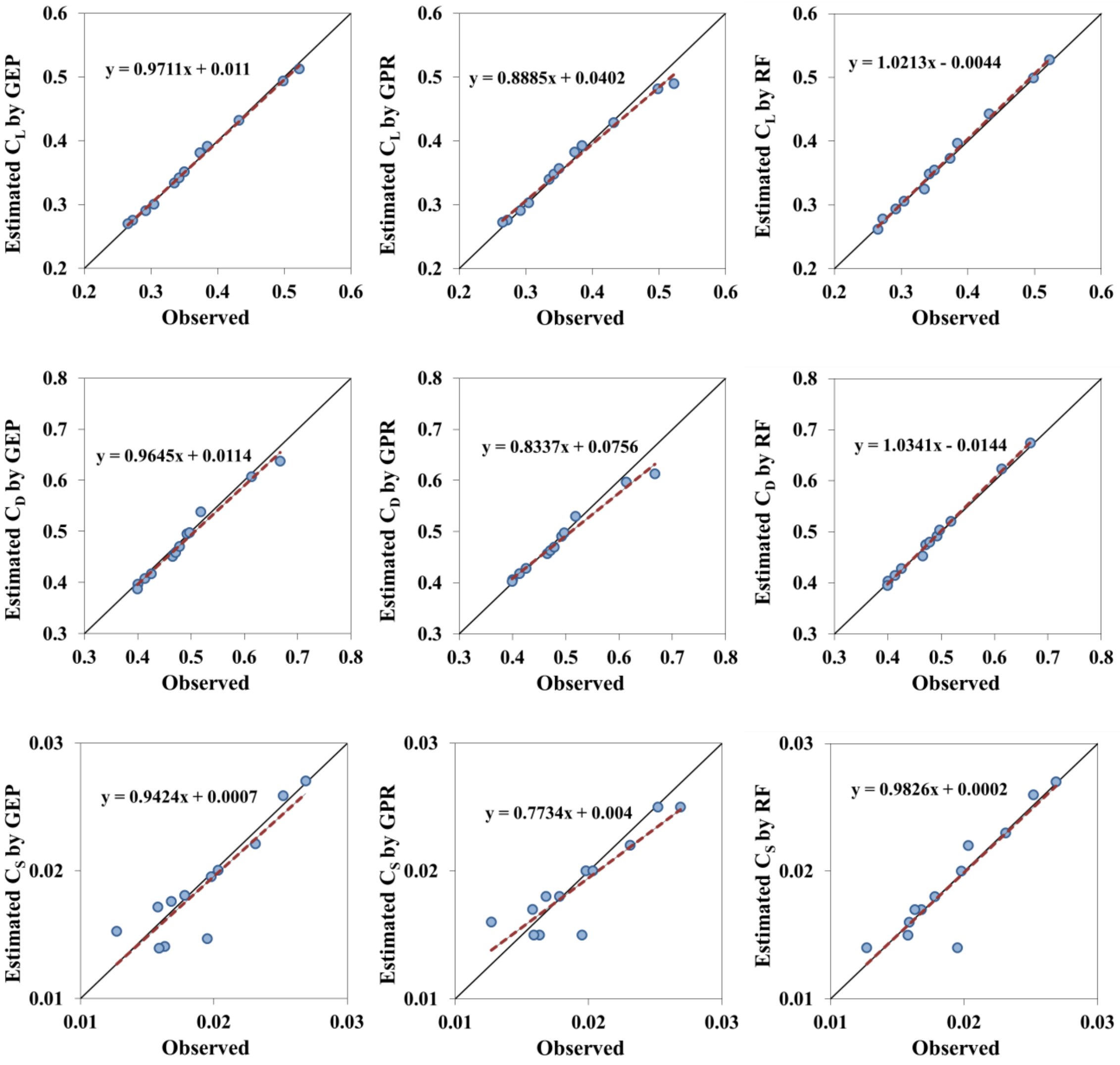

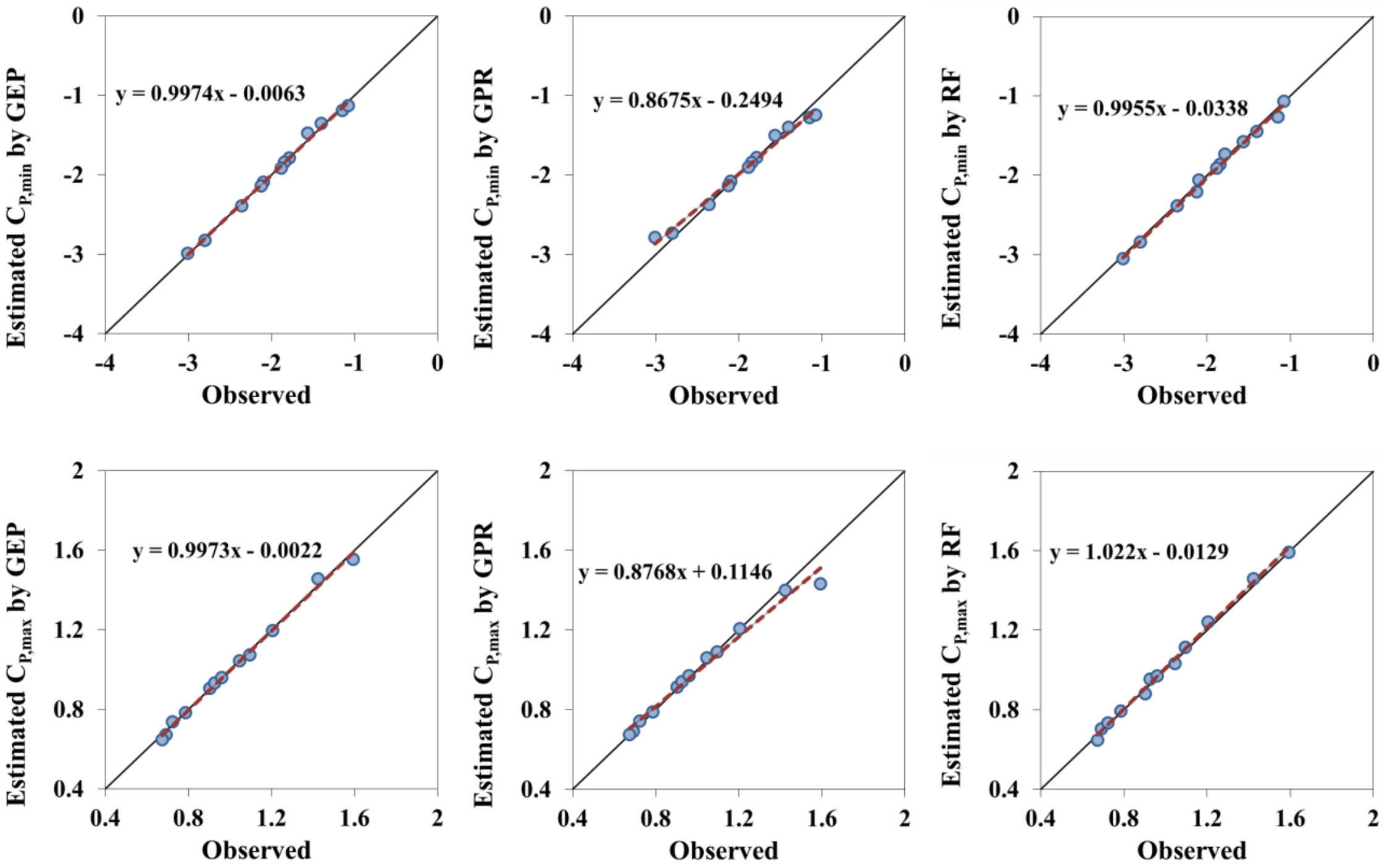

Figure 25. The scatter plots of observed and estimated coefficients (Free stream Velocity).

Despite the lower accuracy of GEP model in the case of aerodynamic coefficients of wind direction and estimating $C_D$, $C_S$, $C_{P,min}$ coefficients in the case of free stream velocity, the produced mathematical formulation of GEP may be practical for the estimation of these aerodynamic coefficients. So, the resulted GEP formulations are presented at Table 9.

Table 9. Resulted GEP Formulae.

| coefficients | | GEP Formulation |
|---|---|---|
| **Wind direction** | $C_L$ | $0.862427\, WD^{0.111111} - \text{ArcTan}\left[e^{2.24744-0.04663\, WD}\, \text{ArcTan}[0.04663 + WD]\right] + \text{ArcTan}\left[\sqrt{\text{ArcTan}\left[1.23153 + 9.54706 \times 10^{9}\, WD\right]}\right]$ |
| | $C_D$ | $0.000153095 \left(e^{\sqrt{WD}} - 9.96466\, WD\right) + 0.00177397\, WD^2 \,(-1.45964 + \text{ArcTan}[WD]) + \text{ArcTan}[\text{ArcTan}[\text{ArcTan}[1.93741 + WD - \text{ArcTan}[WD]]]]^2$ |
| | $C_S$ | $0.0271142\left(WD^{0.333333}+WD\right)-\text{ArcTan}\left[\left(9.78256^{WD}\right)^{0.333333}\right]+\text{ArcTan}[\text{ArcTan}[1.7171+WD]]+\text{Log}\left[\sqrt{WD+\text{Cos}[6.47513\,WD]}\right]$ |
| | $C_{P,min}$ | $-3.37701+e^{0.599632-WD-\text{Cos}[WD]}-\text{ArcTan}[1.36042\,\text{ArcTan}[WD]-\text{Cos}[0.213538\,WD]]+1.27333\,WD^{0.333333}\,\text{Sin}\left[(1.46347+WD)^{0.333333}\right]$ |
| | $C_{P,max}$ | $\left(WD+1.25992\,WD^{0.333333}\,\text{ArcTan}[WD]\right)^{0.333333}+0.237216\,\text{Cos}[4.21558+6.46896\,WD]+\text{Cos}\left[(3.12224\,WD+\text{Sin}[WD])^{0.333333}\right]$ |
| **Free stream Velocity** | $C_L$ | $\frac{2\,FSV}{(5.09753-FSV)(-9.2851+FSV)}+\frac{\text{Cos}[245.28\,FSV]}{-2.60642+1.38367\,FSV}+\text{Sin}\left[e^{-5.07803+\left(FSV^{0.333333}+FSV\right)^{0.333333}}\right]$ |
| | $C_D$ | $(1.23441+FSV)(4.16748+FSV)\,\text{Log}\left[\frac{FSV}{\sqrt{1+FSV^2}}\right]+\left(-0.27922+\sqrt{FSV+\text{Sin}[FSV]}\right)^{0.333333}+\text{Sin}\left[\text{Log}\left[\sqrt{FSV^{2.28339}}+\text{Cos}[0.281435\,FSV]\right]\right]$ |
| | $C_S$ | $-\frac{1.51736+\text{ArcTan}[1.05795+FSV]}{\sqrt{1+(1.51736+\text{ArcTan}[1.05795+FSV])^2}}+\frac{0.00287481+\text{Log}[-4.89496+FSV]}{\sqrt{1+(0.00287481+\text{Log}[-4.89496+FSV])^2}}+\frac{\text{Sin}\left[\sqrt{2}\sqrt{FSV}\right]}{150.385+FSV}$ |
| | $C_{P,min}$ | $-\text{ArcTan}[FSV]+\text{Cos}[1.44681\,\text{ArcTan}[9.9704-FSV]]+\text{Sin}\left[\text{Cos}\left[\sqrt{1.95233+FSV}\right]\right]+\text{Sin}\left[\text{Sin}\left[\sqrt{FSV}\right]-0.0170305\,\text{Sin}[FSV]\right]$ |
| | $C_{P,max}$ | $\text{ArcTan}\left[\text{Log}\left[(5.14609+FSV+\text{Sin}[FSV])^2\right]\right]+0.157244^{-\text{ArcTan}[FSV]}\,FSV^{-\text{ArcTan}[FSV]}\,\text{Cos}\left[2.49701\sqrt{FSV}\right]+\text{Sin}\left[0.87008\,(FSV\,\text{Log}[FSV])^{0.333333}\right]$ |

*WD*: *wind direction* *FSV*: *free stream velocity*

## Conclusion

The basic objective of the present numerical investigation is to analyze the air flow around a high-speed train model. At the first section of the research, a generic high-speed train model is utilized to predict the time-averaged three-dimensional flow structure, turbulence quantities and the aerodynamic forces (as lift, drag, side and pressure coefficients) at different wind direction $\theta$ = 0°, 30°, 45° and 60° and constant velocity magnitude of the free stream with $Re = 2.6 \times 10^6$. The Reynolds Navier-Stokes (RANS) equations combined with the SST $k$-$\omega$ turbulence model are applied to solve incompressible turbulent air flow around the high-speed train. In the following, the influences of velocity (50, 60, 70, 80 and 90 m/s) and the related Reynolds number changes on the flow and aerodynamic key parameters are compared. Also, more detailed results are visible as:

- The flow direction angle has pronounced influence on the three-dimensional flow structure around the model.
- The pressure coefficient  enhances with increasing of wind direction angle.
- The curvy nose of the generic train model has considerable influences on determining the vortex and its turbulent nature at leeward region.
- The distributions of the pressure coefficient are affected by wind direction angle.
- The pressure coefficient have higher magnitude near the nose of the model.

At the second section of the article, GEP, GPR RF methods are used for prediction of the lift, drag and side forces and also minimum and maximum pressure coefficients for wind directions (for $\theta$ = 0˚ to 60˚) and velocity (for 50 m/s to 90 m/s), generally. Due to this methods, the above parameters for all mentioned wind direction and velocity are predicted, simultaneously. Obtained results indicated that RF model performed the ayrodynamic parameters more precisely than GPR and GEP in most cases. In other words, RF provided the superior prredcitions of ayrodynamic parameters in all wind direction parameters and most of free stream velocity parameters. Hence,

RF model may be implemented with high degree of accuracy for predicting ayrodynamic parameters.

## Conflict of interest

The authors declare no conflict of interest.

## Aknowledgment

We acknowledge the support of the German Research Foundation (DFG) and the Bauhaus-Universität Weimar within the Open-Access Publishing Programme. Furthermore, we acknowledge the financial support of this work by the Hungarian State and the European Union under the EFOP-3.6.1-16-2016-00010 project.